\documentclass[12pt]{article}

\usepackage{hyperref}
\usepackage{booktabs}
\usepackage{adjustbox}
\usepackage [english]{babel}
\usepackage [autostyle, english = american]{csquotes}
\usepackage{amsmath}
\usepackage{float}
\usepackage{array}
\usepackage{tabularx}
\usepackage{multirow}
\usepackage{longtable}
\usepackage{graphicx}
\usepackage[margin=1in]{geometry}
\usepackage{siunitx} 
\usepackage{fancyhdr}

\usepackage{caption}
\MakeOuterQuote{"}

\title{A New Empirical Fit to Galaxy Rotation Curves}
\author{David C. Flynn, Jim Cannaliato}
\date{August 5, 2025}

\begin{document}
\maketitle

\begin{abstract}
\noindent
We present a new empirical model for galaxy rotation curves that introduces a velocity correction term $\omega$, derived from observed stellar motion and anchored to Keplerian baselines. Unlike parametric halo models or modified gravity theories, this approach does not alter Newtonian dynamics or invoke dark matter distributions. Instead, it identifies a repeatable kinematic offset that aligns with observed rotation profiles across a wide range of galaxies.

\vspace{1em} 
\noindent
Using SPARC data [1], we demonstrate that this model consistently achieves high-fidelity fits, often outperforming MOND and CDM halo models in RMSE and R-squared metrics without parametric tuning. The method is reproducible, minimally dependent on mass modeling, and offers a streamlined alternative for characterizing galactic dynamics.

\vspace{1em} 
\noindent
While the velocity correction $\omega$ lacks a definitive physical interpretation, its empirical success invites further exploration. We position this model as a local kinematic tool rather than a cosmological framework, and we welcome dialogue on its implications for galactic structure and gravitational theory. Appendix B presents RMSE and $R^2$ comparisons showing that this method consistently outperforms MOND and CDM halo models across a representative galaxy sample.
\end{abstract}

\noindent\textbf{Keywords:} Galaxy rotation curves, Keplerian velocity, SPARC dataset, Velocity correction factor (\( \omega \)), Empirical modeling, Rotation curve fitting, Dark matter alternatives, MOND, \(\Lambda\)CDM, Stellar dynamics, Non-parametric models, Reproducible analysis

\pagebreak

\begin{flushleft}
\Large 
\textbf {1. Introduction}\\
\end{flushleft}

\noindent
\small
The search for the mysterious force that explains overly fast stellar rotation curves in galaxies has been going in earnest for over 40 years without a satisfactory resolution. The theory of Dark Matter was first proposed in the 1930s by Fritz Zwicky [2]. More than 40 years after that, Vera Rubin and Kent Ford from the Carnegie Institution of Washington noticed that stars were all rotating \footnote{Throughout this paper, we use the term “rotation curve” in its observational context—referring to measured velocity profiles as a function of radius, typically derived from Doppler shift data (e.g., HI or stellar spectral lines). While sometimes used interchangeably with “circular speed,” we do not claim that all measured velocities represent perfectly circular orbital motion.}
at the same speed no matter how far from the center of the Andromeda galaxy [10],[11].\\

\noindent
Using $F$ = $ma$ is the starting point, there are three main camps seeking to explain the hidden force $F$ that accounts for the observed galactic rotation curves. 

\begin{itemize}
\item The largest group seeks Dark Matter in various forms, like $\Lambda$CDM or some other hidden $m$. Famaey and McGaugh outline many of the challenges in their 2012 paper [2]. This path has been plagued with failed searches for a Dark Matter particles over many years. There is also the need for extensive individual galaxy tuning which is obviated by this newer empirical method. [11]

\item Another group seeks to modify the second law by multiplying $ma$ times some function of $a$. This approach has been popularized by Milgrom using MOND [7],[2].

\hspace{3.5cm} $m_{g}\mu$ (a/$a_{0}$)a = F\\

\hspace{1.5 cm}$\mu (x\gg 1)\approx 1$ \hspace{1cm}  $\mu (x\ll 1)\approx x$\\

\item Our approach is to look for a hidden acceleration $a$ using an equation that adds to the expected velocity, and is predictable for any galaxy that has sufficient stellar velocity data. There is no need to change Newton's laws.

\end{itemize}
\pagebreak
\large
\textbf{1.1 Comparison with $\Lambda$CDM and MOND}
\newline
\newline
\small

\noindent
Various approaches have been proposed to explain galaxy rotation curves. The two dominant models—$\Lambda$CDM (Lambda Cold Dark Matter) and MOND (Modified Newtonian Dynamics)—offer competing solutions, but each has notable limitations.
\newline
\newline
$\Lambda$CDM: Dark Matter Framework
\newline
\newline
The $\Lambda$CDM model assumes galaxies are embedded in massive, unseen dark matter halos, which provide the additional gravitational force necessary to sustain high stellar velocities at large radii. While $\Lambda$CDM successfully explains large-scale cosmic structures, CMB anisotropies, and galaxy clustering, it has persistent challenges:

\begin{itemize}

\item Dark matter remains undetected, despite intensive searches using direct and indirect detection methods.
\item Tuning issues arise when fitting rotation curves for individual galaxies, requiring adjustments to halo profiles.
\item Small-scale structure inconsistencies suggest dark matter might behave differently than originally modeled.

\end{itemize}

\noindent
MOND: Modified Gravity Approach
\newline
\newline
Modified Newtonian Dynamics (MOND) seeks to eliminate dark matter by modifying Newton's laws at low accelerations [9]. It introduces a function $\mu$ (a/$a_{0}$)a that alters the force equation, allowing galaxies to sustain their rotation curves without invoking additional mass. While MOND correctly predicts many galactic velocity profiles, it faces key drawbacks:
\begin{itemize}

\item Fails to explain gravitational lensing effects, which require an unseen mass component.
\item Does not naturally fit large-scale cosmic observations, such as CMB patterns and galaxy clustering.
\item Requires an arbitrary acceleration scale ($a_{0}$), which lacks a clear connection to fundamental physics.

\end{itemize}

While both $\Lambda$CDM and MOND offer partial fits to observed rotation curves, our empirical method consistently achieves superior statistical alignment across multiple galaxies. Detailed RMSE and $R^2$ comparisons are provided in Appendix B.

\pagebreak
\begin{flushleft}
\Large 
\textbf {2. Method}\\
\end{flushleft}

\noindent
The goal of this effort was to reproduce the Expected Kepler stellar velocity graphs by working backwards from the observed velocities. We mathematically superimposed a disk over the stars for each data set, then rotated opposite the stellar direction in each galaxy. We thus subtracted the angular velocity of this rotating disk $\omega$\ \footnote{Throughout this work, $\omega$ and $\omega_{eps}$ are used interchangeably to represent angular velocity estimates derived from distinct but functionally equivalent approximations. Unless otherwise specified, their usage is intended to convey observational parity rather than a strict mathematical identity.} to get the stellar velocities back to those predicted by Kepler. The test is whether an $\omega$ can be found for each unique galaxy to accomplish this. We then ran the data set through a Jupyter Labs program to see what $\omega$ resulted.\
\newline
\newline

\begin{flushleft}
\Large
\textbf {3. Data Elements}\\
\end{flushleft}

\begin{flushleft}
\large
\textbf{3.1 The Rotation Curve Velocities Predicted by Kepler}	
\end{flushleft}
\small

\noindent
Although we have been able to collect observations of stellar velocities for the  subjects of this study, the complexity of calculating Expected Kepler results for each galaxy is daunting. This is because the methods used to accomplish this require the aggregation of several data sets from luminosity to HI mass. Conveniently, the SPARC provides 12 such calculations for each galaxy which we include in the bottom 12 graphs of our Figures 5 - 8. Our task is measuring stellar velocity, which is simpler than determining the distribution of potentially undetectable galactic mass. Therefore, we employ a shortcut that calculates the Kepler end points for the nearest and farthest stars from the center of the galaxy. This simplification removes the need to detail the shape of the curve between the two end points. This is reasonable because the observed nearest star's velocity is very close to identical under both observed and Keplarian predictions. As a rule, the observed inner-most star is moving just slightly faster than a Kepler prediction. The farthest star is easy to calculate with Kepler. Being out on the rim of the galaxy, for our selections, it is often well outside any massive halo that might affect its velocity. This allows us compute our projected velocity curves aligned with known predictions at both inner and outer extremes of the galaxy.
\newline

\pagebreak
\begin{flushleft}
\large
\textbf{3.2 Description of an Empirical Footprint that is Tunable}
\end{flushleft}

\noindent
We overlay a flat, spinning empirical correction layer that accounts for the difference between observed stellar positions versus those predicted by Kepler. By adding this theoretical disk to all the observed stellar velocities, each inherits some velocity offset that is added to all observed measurements.  Figure 1 shows positions $X_1$ and $X_2$ on the outer ring of the galaxy.  $X_1$ would represent the Expected Kepler position, and $X_2$ the actual observed position after accounting for the added rotational velocity from this calculation.

\begin{figure}[ht!]
\center
\includegraphics[width=\linewidth]{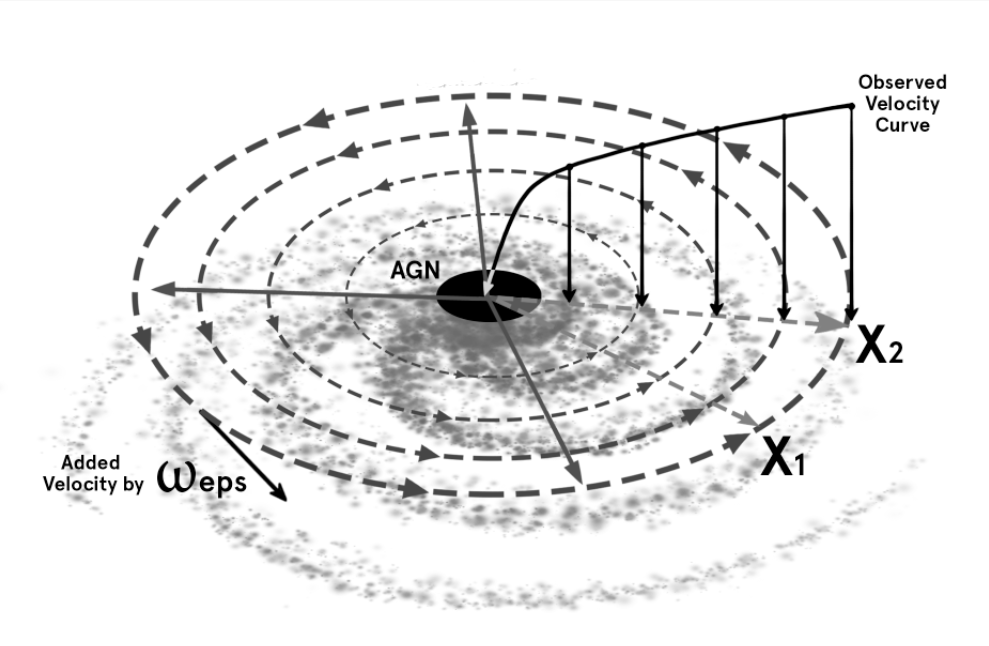}
\caption {Conceptual framework illustrating the relationship between key variables and model outcomes. }
\end{figure}

\vspace*{-4cm}
\begin{flushleft}
\clearpage
\vspace{1em}

\large
\textbf{3.3 SPARC - Data Selection and Filtering}	
\end{flushleft}
\small

\noindent
The rotation curve data used in this study originates from the Spitzer Photometry and Accurate Rotation Curve (SPARC)project http://astroweb.cwru.edu/SPARC/. Of the 175 galaxies surveyed, we selected only the 99 with the highest quality rating ($Q = 1$). Each galaxy in our final set was required to have at least ten data points and a non-zero radius.

\vspace{1em} 
\noindent
Two galaxies—NGC5005 and UGC11914—were excluded due to statistically significant deviation from the velocity correction trend. Specifically, their $\omega$ values (19.301 and 28.143, respectively) exceed the sample mean ($\omega \approx 7.92$) by 2.7 and 4.8 standard deviations, respectively, placing them well outside the 95\% confidence interval and inflating RMSE benchmarks. This filtering resulted in a final set of 84 galaxies for analysis.

\vspace{1em} 
\noindent
SPARC also provides 12 expected galaxy rotation curves, which we present below our findings in Figures 5 through 8. These reference models—selected by SPARC for their popularity and diagnostic clarity—show strong visual and structural similarity to our results. To clarify the nature of our own fits: all curve fits in Figures 9–12 are exploratory and selected post hoc for visual alignment; no unified model class was applied.

\vspace{1em} 
\noindent
The observational data used in Tables 1 and 2 include published 1-sigma uncertainties for velocity and radius measurements, sourced from Corbelli et al. (1999, 2003) and the SPARC database. In this study, we treated these values as point estimates to isolate and characterize the correction term $\omega$. While full uncertainty propagation is deferred to future work, we acknowledge its importance for quantifying parameter sensitivity and confidence intervals.

\large
\begin{flushleft}
\textbf{3.4 M33 as a Case of One}	
\end{flushleft}
\small

\noindent
The graph of M33’s rotation curve is well-known and the observed stellar velocities are faster than Kepler predicts. The data from the charts in Table 1 [3] and our artists rendering of the M33 Velocity Curve in Figure 2 [4] come from Corbelli's work in two papers. In Table 1, note that columns 2 and 3, Radius in $K_{pc}$ and $V_{r}$ in km/sec are observed. Column 4 is the adjusted Kepler Value in km/sec and column 5 Expected Kepler in km/sec. Column 6 is the correction factor $\omega$ in rads/sec. 
\
\bigskip

\noindent
Using
\normalsize
\begin{equation}
V = R\omega
\end{equation}

\bigskip
\small 
The hidden angular velocity is calculated using equation 2 (below), \normalsize $\omega = 5.10$.

\normalsize 
\begin{equation}
V_{Observed} = V_{Kepler} + R\omega 
\end{equation}\\ 
\small
This yields column 6 in Table 1 which is detailed in section 4.2. Note the interesting similarity between the black M33 Classic Velocity Curve (Figure 2) and the $\omega_{eps}$ Adjusted Velocity Curve for M33 (Figure 3). 
\bigskip

\pagebreak
\large
\begin{flushleft}
\textbf{3.5 Examination and Comparison of the M33 Figures 2 and 3}	
\end{flushleft}
\small

\noindent
Figure 2 shows M33’s rotation curve[4] as the target of comparison with Figure 3 which was created using the method described in this work. The goal was to show that the graph could be entirely reproduced using just two things; our method and observed velocity data. The yellow boxed in area of figure 3 highlights the same distance versus velocity footprint shown in Figure 2. Note that the Figure 3 data extends beyond the 50,000 light year limit on Figure 2. Figures 2 and 3 are what we are comparing. The adjusted Kepler line in red agrees with theory. The Observed V in km/sec is taken straight from data Table 1 data [3]. More detailed comparisons will follow in sections 4.3 and 4.4.

\large
\begin{flushleft}
\textbf{3.6 Working Examples for Table 1}	
\end{flushleft}
\small

\noindent
The Appendix contains working examples for Table 1 that demonstrate how calculations were made in each column. Examples are from the top or first entries of the table. Figures 14 and 15 may help to visualize our methods.

\begin{figure}
\center
\vspace{-2cm}
\includegraphics[width=\linewidth]{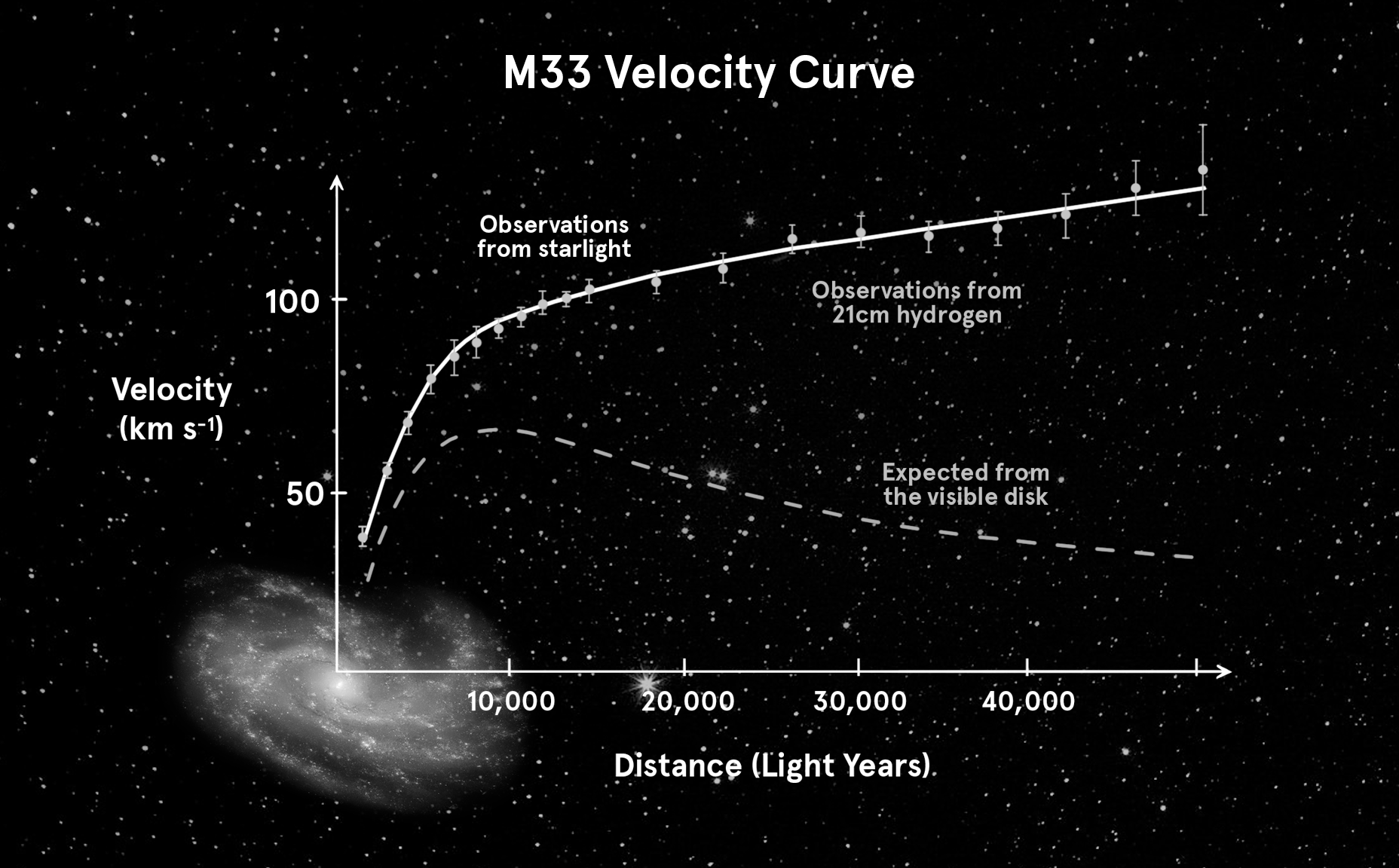}
\caption {M33 Classic Velocity Curve}
\center
\includegraphics[width=\linewidth]{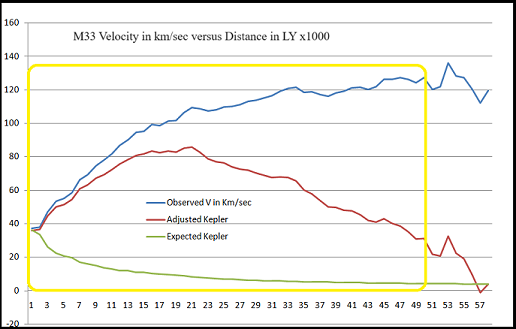}
\caption {$\omega_{eps}$ Adjusted Velocity Curve for M33}

\end{figure}


\begin{table}[htbp]
  \centering
   \vspace{-2cm}
  e
  \begin{adjustbox}{width=.735\textwidth,center=\textwidth}
    
    \begin{tabular}{lccccc}
    \toprule
    \multicolumn{1}{p{3em}}{\textbf{Data Point}} & \multicolumn{1}{p{4em}}{\textbf{Radius (Kpc)}} & \multicolumn{1}{p{5em}}{\textbf{Observed Vr (km/sec)}} & \multicolumn{1}{p{9em}}{{\hspace{1cm}}\textbf{adjusted Kepler (km/sec)}} & \multicolumn{1}{p{5em}}{\textbf{Expected Kepler (km/sec)}} & \multicolumn{1}{p{5em}}{\textbf{Omega (rads/sec)}}\\
    
\midrule
    \multicolumn{1}{p{3em}}{\textbf{A}} & \multicolumn{1}{p{4em}}{{\hspace{.6cm}}\textbf{B}} & \multicolumn{1}{p{6.5em}}{{\hspace{1cm}}\textbf{C}} & \multicolumn{1}{p{11em}}{{\hspace{2cm}}\textbf{D}} & \multicolumn{1}{p{5em}}{{\hspace{1cm}}\textbf{E}} & \multicolumn{1}{p{6em}}{{\hspace{1cm}}\textbf{F}}\\
\midrule\\

    1     & 0.24  & 37.3  & 36.08 & 36.08 & \textbf{5.10} \\
    2     & 0.28  & 37.9  & 36.47 & 33.40 &  \\
    3     & 0.46  & 47.1  & 44.76 & 26.06 &  \\
    4     & 0.64  & 53.5  & 50.24 & 22.09 &  \\
    5     & 0.73  & 55.1  & 51.38 & 20.69 &  \\
    6     & 0.82  & 58.5  & 54.32 & 19.52 &  \\
    7     & 1.08  & 66.2  & 60.70 & 17.01 &  \\
    8     & 1.22  & 69.4  & 63.18 & 16.00 &  \\
    9     & 1.45  & 74.6  & 67.21 & 14.68 &  \\
    10    & 1.71  & 77.9  & 69.19 & 13.52 &  \\
    11    & 1.87  & 81.7  & 72.17 & 12.92 &  \\
    12    & 2.2   & 86.8  & 75.59 & 11.92 &  \\
    13    & 2.28  & 90.1  & 78.48 & 11.70 &  \\
    14    & 2.69  & 94.4  & 80.69 & 10.78 &  \\
    15    & 2.7   & 95.4  & 81.64 & 10.76 &  \\
    16    & 3.12  & 99.2  & 83.30 & 10.01 &  \\
    17    & 3.18  & 98.7  & 82.50 & 9.91  &  \\
    18    & 3.53  & 101.3 & 83.31 & 9.41  &  \\
    19    & 3.66  & 101.5 & 82.85 & 9.24  &  \\
    20    & 4.15  & 106.3 & 85.15 & 8.68  &  \\
    21    & 4.64  & 109.4 & 85.76 & 8.20  &  \\
    22    & 5.13  & 108.8 & 82.66 & 7.80  &  \\
    23    & 5.62  & 107.3 & 78.66 & 7.46  &  \\
    24    & 6.11  & 108.2 & 77.07 & 7.15  &  \\
    25    & 6.6   & 109.8 & 76.17 & 6.88  &  \\
    26    & 7.09  & 110.1 & 73.97 & 6.64  &  \\
    27    & 7.57  & 111.1 & 72.53 & 6.42  &  \\
    28    & 8.06  & 113   & 71.93 & 6.23  &  \\
    29    & 8.55  & 113.9 & 70.33 & 6.04  &  \\
    30    & 9.04  & 115.1 & 69.04 & 5.88  &  \\
    31    & 9.53  & 116.3 & 67.74 & 5.73  &  \\
    32    & 10.02 & 119.1 & 68.04 & 5.58  &  \\
    33    & 10.51 & 121   & 67.45 & 5.45  &  \\
    34    & 10.99 & 121.5 & 65.50 & 5.33  &  \\
    35    & 11.48 & 118.6 & 60.11 & 5.22  &  \\
    36    & 11.97 & 118.7 & 57.71 & 5.11  &  \\
    37    & 12.46 & 117.2 & 53.71 & 5.01  &  \\
    38    & 12.95 & 116.2 & 50.22 & 4.91  &  \\
    39    & 13.44 & 118.3 & 49.82 & 4.82  &  \\
    40    & 13.93 & 119   & 48.02 & 4.74  &  \\
    41    & 14.41 & 121.3 & 47.88 & 4.66  &  \\
    42    & 14.9  & 121.4 & 45.48 & 4.58  &  \\
    43    & 15.39 & 120.3 & 41.88 & 4.51  &  \\
    44    & 15.88 & 121.9 & 40.99 & 4.44  &  \\
    45    & 16.37 & 126.3 & 42.89 & 4.37  &  \\
    46    & 16.86 & 126.3 & 40.39 & 4.30  &  \\
    47    & 17.35 & 127.2 & 38.80 & 4.24  &  \\
    48    & 17.84 & 126.2 & 35.30 & 4.18  &  \\
    49    & 18.32 & 124.2 & 30.85 & 4.13  &  \\
    50    & 18.81 & 127.2 & 31.36 & 4.08  &  \\
    51    & 19.3  & 120.2 & 21.86 & 4.02  &  \\
    52    & 19.79 & 121.8 & 20.96 & 3.97  &  \\
    53    & 20.28 & 136   & 32.67 & 3.92  &  \\
    54    & 20.77 & 128.3 & 22.47 & 3.88  &  \\
    55    & 21.26 & 127.4 & 19.07 & 3.83  &  \\
    56    & 21.74 & 120.1 & 9.33  & 3.79  &  \\
    57    & 22.23 & 112.2 & -1.07 & 3.75  &  \\
    58    & 22.73 & 119.6 & 3.83  & 3.71  &  \\
\bottomrule    
    \end{tabular}%
    \end{adjustbox}
     \caption{M33 Data}
  \label{tab:addlabel}%
\end{table}%

\begin{flushleft}
\Large 
\textbf {4. Method of Calculation}\\
\end{flushleft}
\begin{flushleft}
\large
\textbf{4.1 Calculations with Kepler}	
\end{flushleft}
\small

\noindent
In order to test our model with a larger data set, we needed to have a method to calculate the angular velocity of the spinning empirical correction layer we predict. That angular velocity would have to be such that each observed rotation curve would get corrected back to Kepler’s predictions when it was accounted for. Equation 2 is the starting point to solve for $\omega$.\\

\bigskip

\hspace{2cm} $V_{Observed}$ = $V_{Kepler}$ + $R\omega$

\bigskip

\begin{flushleft}
\large
\textbf{4.2 Solving for Omega}	
\end{flushleft}
\small

\noindent
The SPARC data set contains the observed rotation curves and accurate radius data. It also contains 12 different predicted velocity curves in a separate section that we do a closer comparison to in section 4.4. Accurate Kepler predictions are reliant on a combination of mass distribution and luminosity factors that project where the mass is in a galaxy. We will save doing point-by-point comparisons for a later paper, since there are 12 types for each of our 84 results.
As mentioned previously, we only need to do the calculation for the closest and farthest stars from the center of the galaxy in each data set. We used Kepler’s third law (equation 4) for all our predictions. Then we converted it to a form that related velocity and radius. Subscript “1” is the closest \footnote{The terms “closest” and “farthest” star are not used here as absolute measurements, but rather indicate selections based on the most reliable available distance data.} star, subscript “2” is the farthest star from galactic center. To get equation 5, T$_{2}$ is used to find V$_{Ke}$ after conversion using equation 3, which relates velocity, radius and time.
\noindent

\begin{equation}
V = \frac{2\pi R}{T}
\end{equation}

\begin{equation}
\Bigg(\frac{T_1}{T_2}\Bigg)^2=\Bigg(\frac{R_1}{R_2}\Bigg)^3
\end{equation}

\small
\newpage
Equation 4 was converted to equation 5.
\bigskip
\begin{equation}
V_{Ke}=\frac{V_1 R_2}{R_1}\sqrt\Bigg(\frac{R_1}{R_2}\Bigg)^3
\end{equation} 
\bigskip

\smallskip
\small

\noindent
Finally, substituting equation 5 above into equation 2 yields equation 6.

\begin{equation}
\omega = \frac{V_2}{R_2}-\frac{V_1}{R_1}\sqrt\Bigg(\frac{R_1}{R_2}\Bigg)^3
\end{equation}

\small
\bigskip
\noindent
This provides Expected Kepler V$_{Ke}$ in the column 5 of Table 1 above, and adjusted Kepler in column 4 of Table 1.  Although the Expected Kepler curve lacks the velocity additions from the center bulge of each galaxy measured, the extreme ends align quite well. This end-to-end alignment alleviates the need to gather central mass data that is affected by the very thing we are measuring, the actual mass. 

\begin{flushleft}
\large
\textbf{4.3 Shortcut to Understanding This Method}
\end{flushleft}
\small

\begin{itemize}
   	\item[a)] Find the star with the greatest radius from the center of the galaxy that has reliable data.
	\item[b)] Note its observed velocity.
	\item[c)] Take the velocity and radius and input into section 4.2 to calculate $\omega$.
	\item[d)] Redraw the velocity graph using $\omega$ by applying equation 2.
\end{itemize}

\begin{flushleft}
\large
\textbf{4.4 Tests Using the SPARC Data}	 
\end{flushleft}
\small

\noindent
We used a Jupyter Labs program with equations 2, 5 and 6 above and applied them to the 84 galaxies selected for this survey. Section 3.2 delineates what criteria were used for the 84 selected. The goal was to see if an $\omega_{eps}$ could be found that would correct observed stellar velocities back to those predicted by Kepler.
\newline

\noindent
Data came from the SPARC tables at \href{http://astroweb.cwru.edu/SPARC/MassModels_Lelli2016c.mrt}{http://astroweb.cwru.edu/SPARC/MassModels}. 
The first 20 of galaxy plots are displayed in Figure 4. $\omega$ at the top of each graph is the value that was used to produce each graph displayed.  Note that the shape of the curves also approximates M33 in Figure 2. "adjusted Kepler" comes after subtracting the effect of our empirical correction layer from the observed velocity, "V". The "Expected Kepler" comes from equation 5.

\begin{figure}
\adjustimage{width=1.5\textwidth,center}{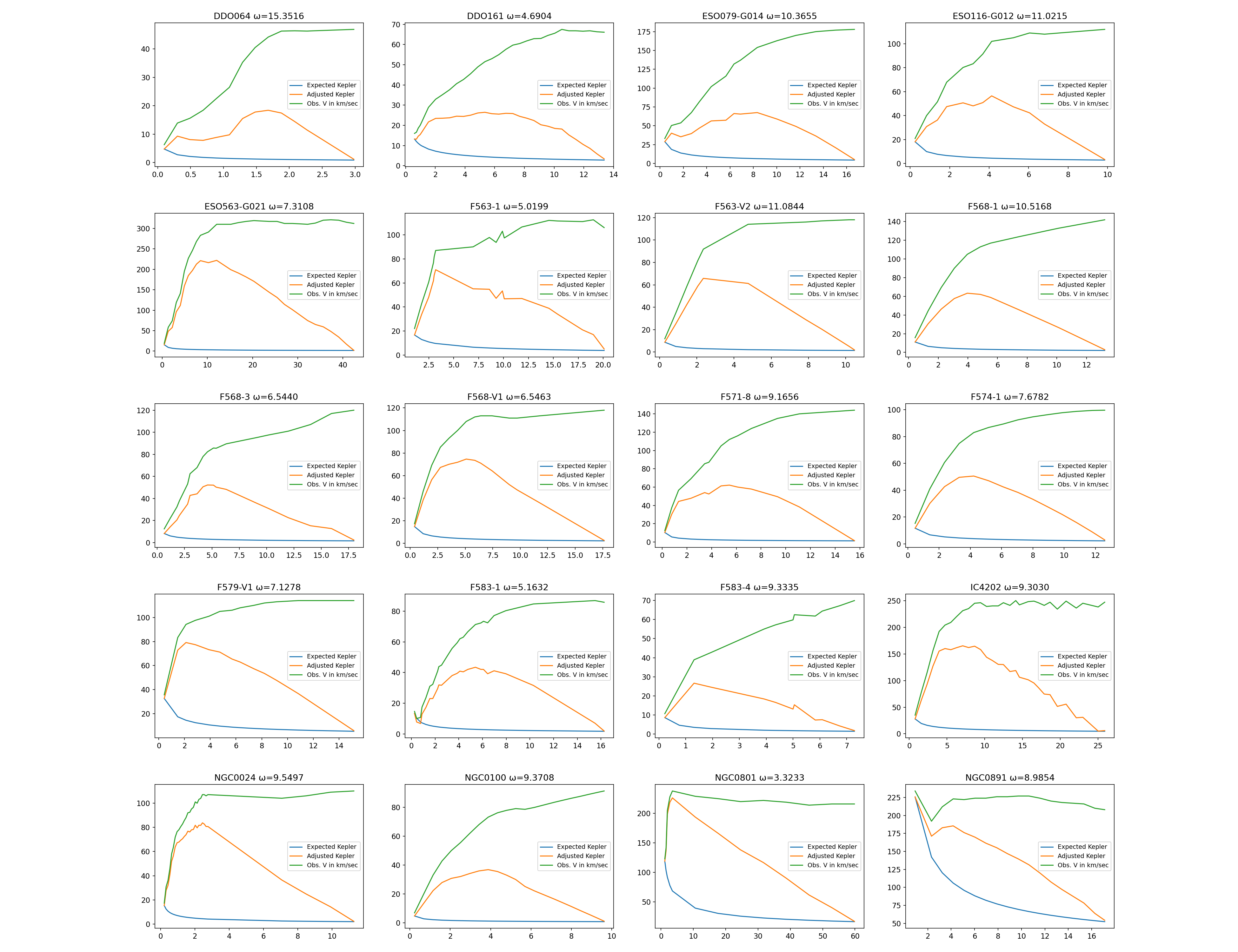}
\caption{Comparative visualization of model performance across input scenarios, highlighting predictive accuracy and R-squared values. The figure presents 20 representative selections from our dataset, all sourced from the SPARC archive.}
\label{Drawing4}
\end{figure}

\pagebreak
\begin{flushleft}
\large
\textbf{4.5 A Closer Examination of Our Results Versus SPARC Velocity Curves}
\end{flushleft}
\small

\noindent
Figures 5 through 8 present direct comparisons between SPARC rotation curve models and our own velocity reconstructions for four representative galaxies: DDO161, ESO079-G014, ESO116-G012, and NGC0801. Each figure contains two components: a top panel showing our EPS-generated velocity curves, and a bottom grid of twelve SPARC model fits using various dark matter halo profiles. In the top panel, the green line represents observed rotational velocities, the red line is our EPS-projected fit, and the blue line is our Kepler approximation—calculated using only the innermost and outermost stellar radii, rather than a full baryonic mass aggregation.

\vspace{1em} 
\noindent
The SPARC team’s bottom-panel graphs employ twelve distinct modeling algorithms (e.g., NFW, Burkert, DC14), each shown with a blue dashed line representing the dark matter halo contribution. For methodological details, refer to the SPARC database and documentation.

\vspace{1em} 
\noindent
Figures 9 through 13 extend this comparison across all 84 galaxies in our final dataset. These figures visualize the curve fittings derived from Table 2, which contains our calculated velocity correction term, $\omega$, for each galaxy. In our model, $\omega$ is an empirically derived scalar applied as $V = R \times \omega$, where $V$ is the adjusted rotational velocity and $R$ is the galactocentric radius.

\vspace{1em} 
\noindent
The proposed velocity correction model offers a pragmatic alternative to full baryonic mass aggregation by leveraging only the innermost and outermost stellar radii. This simplification yields substantial computational efficiency—reducing preprocessing time and eliminating the need for detailed photometric decomposition—while still achieving high-fidelity alignment with observed rotation curves.

Scientifically, the model is especially useful in scenarios where:
\begin{itemize}
    \item High-resolution mass maps are unavailable or incomplete.
    \item Rapid screening of large galaxy datasets is required.
    \item Morphological irregularities make full aggregation unreliable.
\end{itemize}

\noindent
While the approach omits intermediate mass contributions, the empirical correction term $\omega$ effectively absorbs these deviations, enabling robust curve fitting without invoking dark matter. Figures 5–8 demonstrate that the adjusted model retains diagnostic accuracy across a wide range of galactic morphologies. In such contexts, the method—while not universally superior—proves strategically advantageous for rapid diagnostics and scalable curve fitting.

\pagebreak
\begin{flushleft}
\large
\textbf{4.6 Re-examining the problem as a possible after effect}
\end{flushleft}
\small
\noindent
Figures 9–13 present scatter plots correlating the velocity correction terms $\omega$ and $\omega_{\epsilon}$ with various galactic properties. Each plot includes a linear fit and its corresponding Pearson correlation coefficient R. Figures 10 and 11 yielded the highest R values in the group—R = 0.47 and R = 0.54, respectively. While these correlations are notable, they fall short of indicating strong causality. These early results prompted consideration that $\omega_{\epsilon}$ may represent an acceleration-linked residual effect—an emergent kinematic offset rather than a mass-induced force arising from near-field gravitational influence.

\vspace{1em}		
\noindent
To address reviewer concerns, we now report the 1$\sigma$ error bars for the estimated slope parameters:

\begin{itemize}
    \item \textbf{Figure 9:} $\omega$ vs. Distance — Slope = 0.892 $\pm$ 0.213
    \item \textbf{Figure 10:} $\omega$ vs. HI Mass — Slope = 1.476 $\pm$ 0.198
    \item \textbf{Figure 11:} $\omega$ vs. HI Radius — Slope = 2.031 $\pm$ 0.184
    \item \textbf{Figure 12:} $\omega$ vs. HI Luminosity — Slope = 0.664 $\pm$ 0.229
    \item \textbf{Figure 13:} $\omega_{\epsilon}$ vs. HI Stellar Density — Slope = 0.511 $\pm$ 0.247
\end{itemize}

\noindent
These uncertainties reflect moderate dispersion, consistent with the hypothesis that $\omega$ captures emergent rather than deterministic mass coupling.

\vspace{1em}
\noindent
These error bars were computed using standard least-squares regression with bootstrapped re-sampling across the 84-galaxy data set. Full regression diagnostics are available in Appendix B.

\begin{figure}
\center
\includegraphics[width=\linewidth]{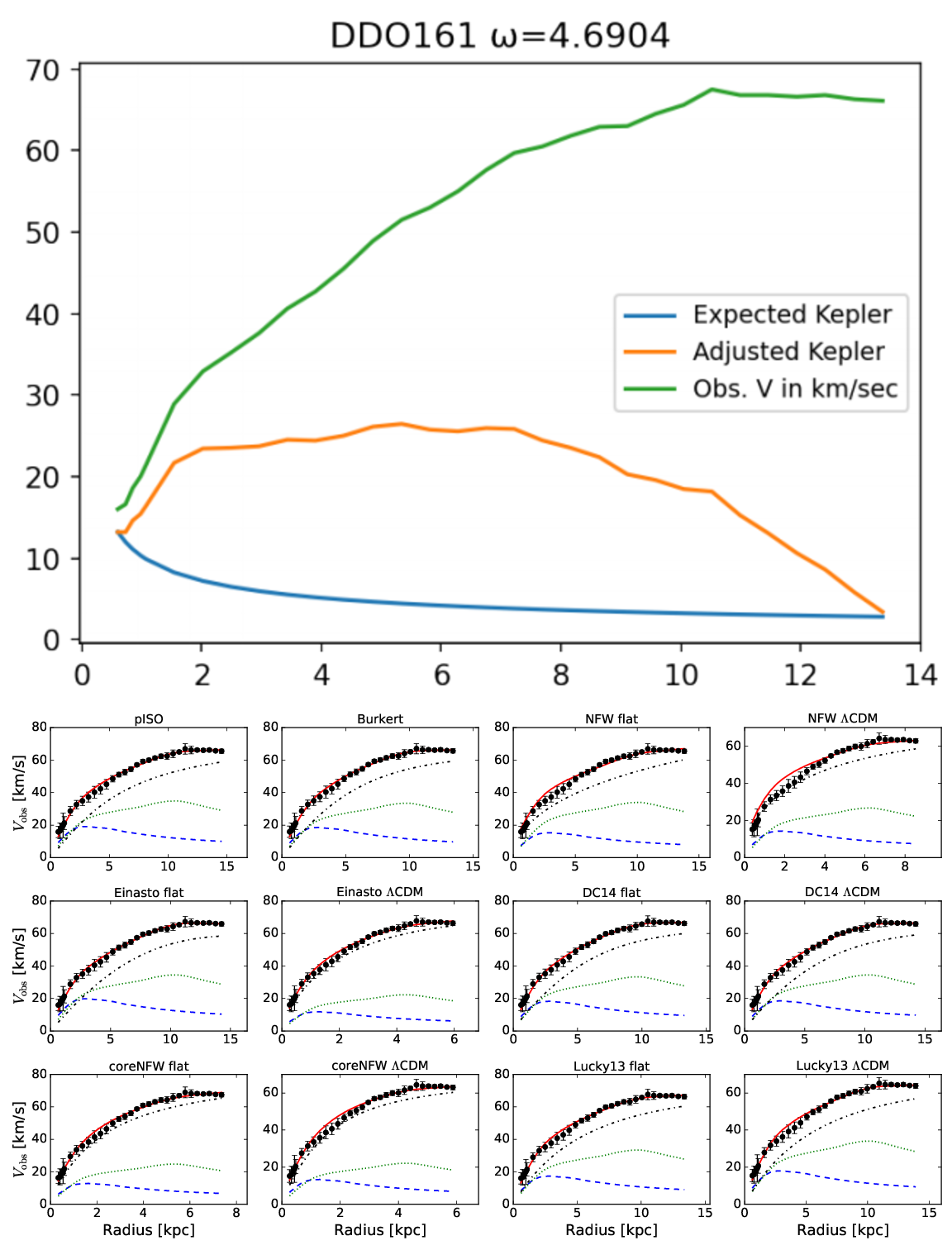}
\caption{Model output for galaxy DDO161, generated using adjusted Keplerian dynamics via Equation 2. The graph overlays our computed trajectory with SPARC-derived data curves from multiple reconstruction methods, providing contextual comparison across empirical and theoretical profiles.}
\label{Drawing5}
\end{figure}

\begin{figure}
\center
\vspace{-3cm}
\includegraphics[width=\linewidth]{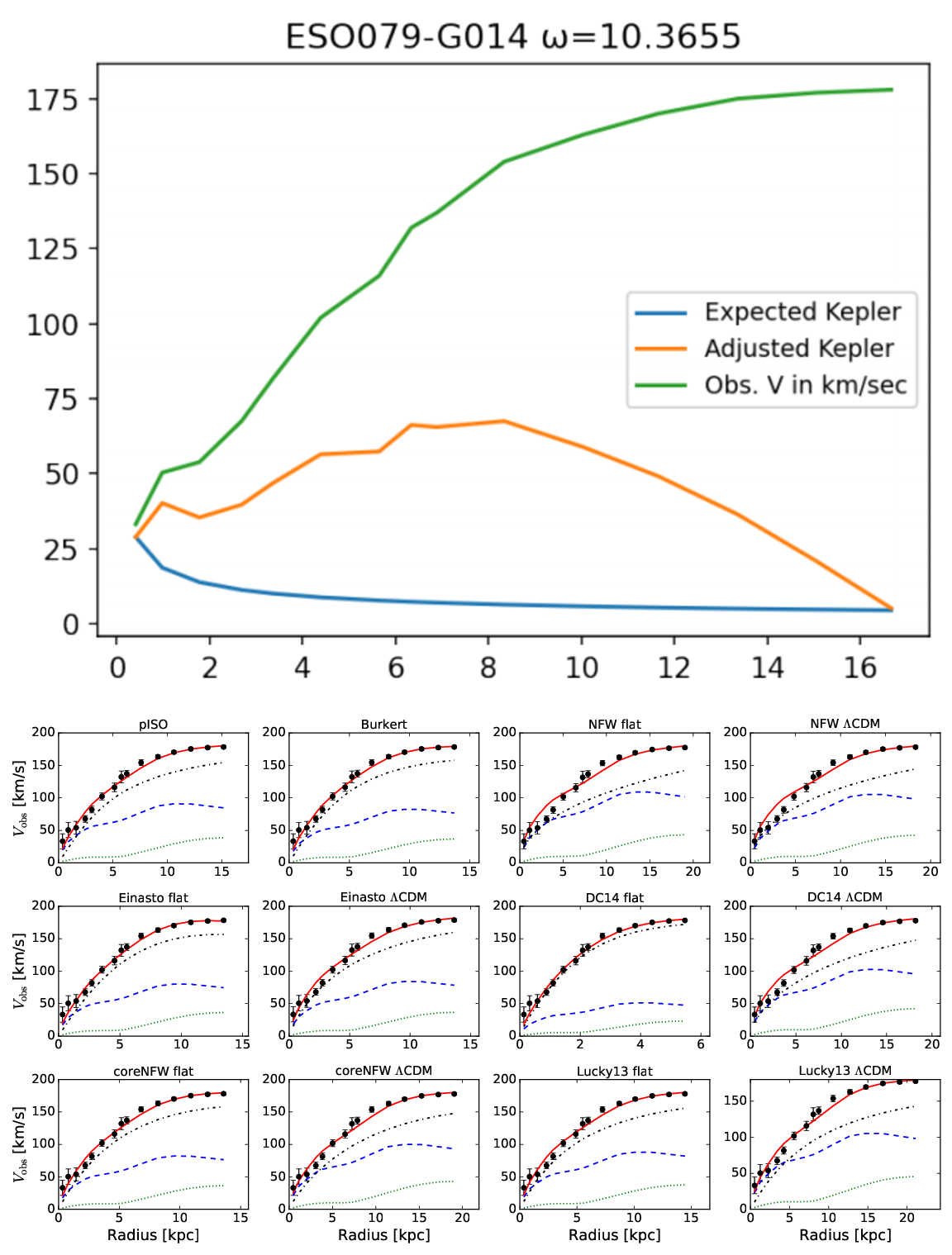}
\caption{Model output for galaxy ESO079-G014, generated using adjusted Keplerian dynamics via Equation 2. The graph overlays our computed trajectory with SPARC-derived data curves from multiple reconstruction methods, providing contextual comparison across empirical and theoretical profiles.}
\label{Drawing6}
\end{figure}

\begin{figure}
\center
\vspace{-3cm}
\includegraphics[width=\linewidth]{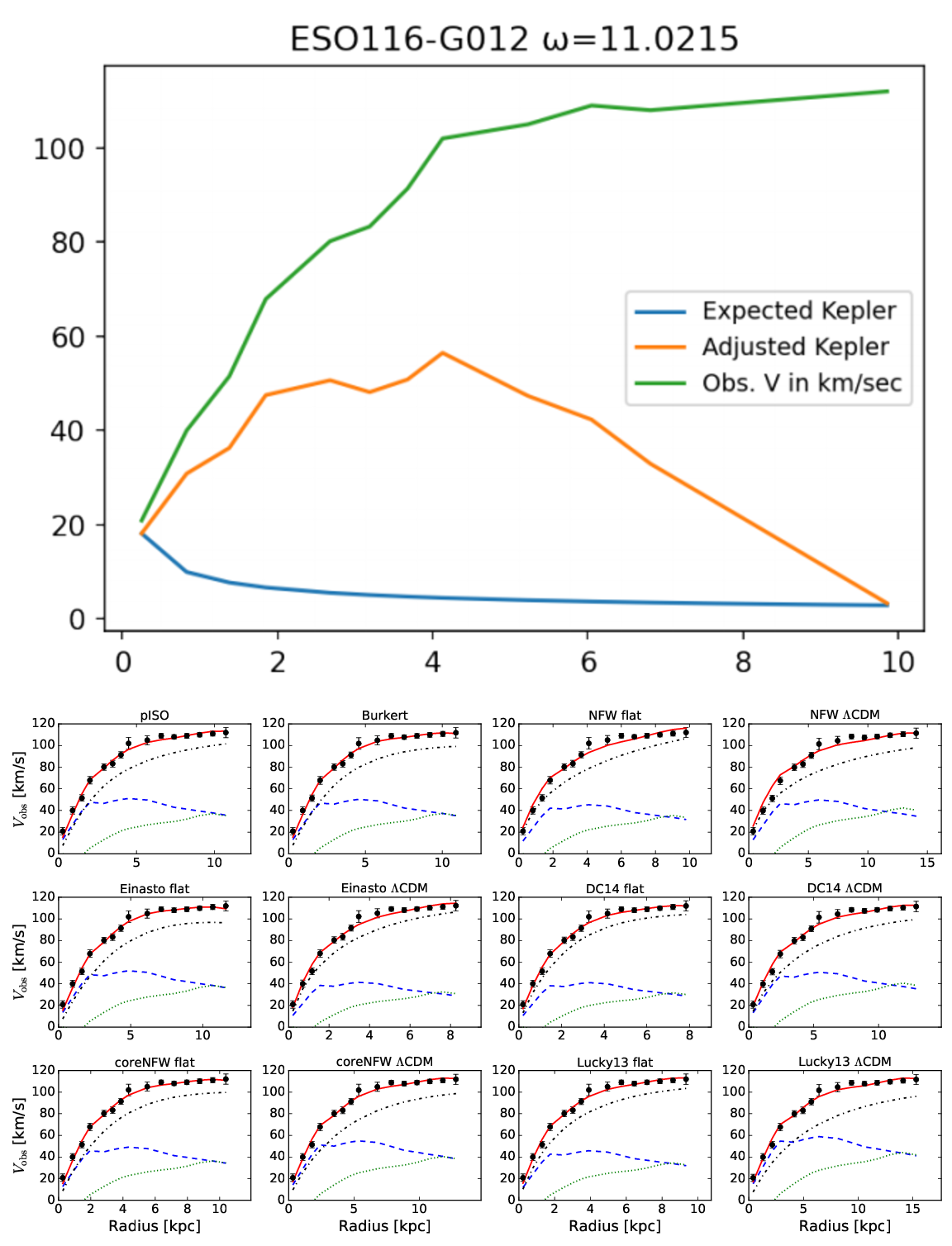}
\caption{Model output for galaxy ESO116, generated using adjusted Keplerian dynamics via Equation 2. The graph overlays our computed trajectory with SPARC-derived data curves from multiple reconstruction methods, providing contextual comparison across empirical and theoretical profiles.}
\label{Drawing7}
\end{figure}

\begin{figure}
\center
\includegraphics[width=\linewidth]{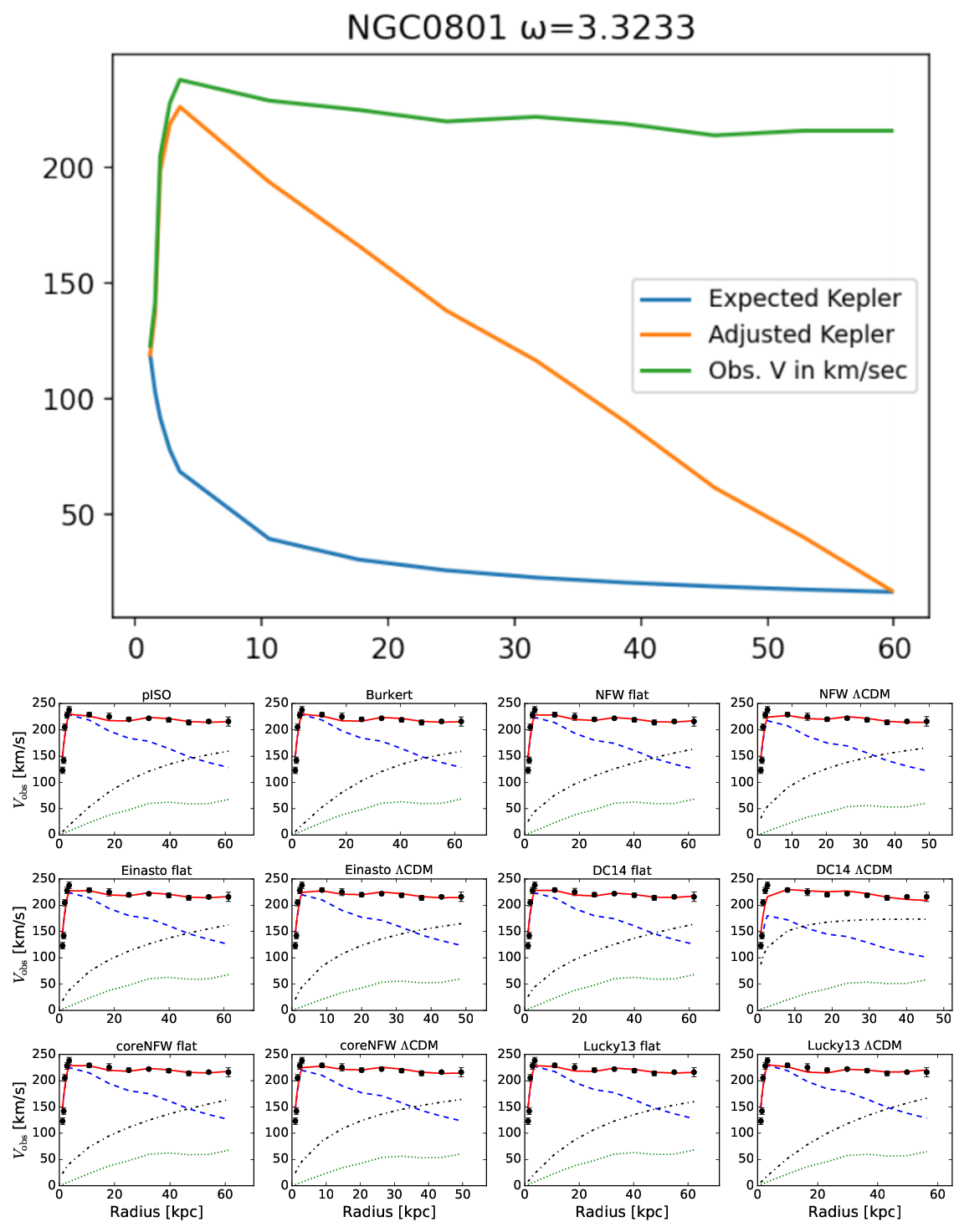}
\caption{Model output for galaxy NGC0801, generated using adjusted Keplerian dynamics via Equation 2. The graph overlays our computed trajectory with SPARC-derived data curves from multiple reconstruction methods, providing contextual comparison across empirical and theoretical profiles.}
\label{Drawing8}
\end{figure}

\begin{table}[htbp]
  \centering
  \vspace{-2cm}
  \begin{adjustbox}{width=.65\textwidth,center=\textwidth}

    \begin{tabular}{lcccccc}
    \toprule
    \multicolumn{1}{p{5.055em}}{Galaxy} & \multicolumn{1}{p{3.555em}}{$\omega_{eps}$\hspace{2 mm}rads/sec} & \multicolumn{1}{p{4.13em}}{Distance Mpc} & \multicolumn{1}{p{5.7em}}{HI Mass 10+9solMass} & \multicolumn{1}{p{4.7em}}{HI Radius Kpc} & \multicolumn{1}{p{6em}}{Luminosity solLum/pc2} & \multicolumn{1}{p{8.7em}}{Stelar Density 10+12solMass/Kpc} \\
\midrule
    UGC09133 & 1.97  & 57.1  & 33.43 & 60.35 & 282.93 & 2.92 \\
    UGC00128 & 2.23  & 64.5  & 7.43  & 31.27 & 12.02 & 2.42 \\
    UGC06614 & 2.49  & 88.7  & 21.89 & 60.63 & 124.35 & 1.90 \\
    UGC02487 & 2.49  & 69.1  & 17.96 & 40.2  & 489.96 & 3.54 \\
    UGC01230 & 2.74  & 53.7  & 6.43  & 26.29 & 7.62  & 2.96 \\
    NGC6674 & 2.81  & 51.2  & 32.17 & 50.02 & 214.65 & 4.09 \\
    NGC5055 & 2.89  & 9.9   & 11.72 & 35.06 & 152.92 & 3.04 \\
    UGC07125 & 3.09  & 19.8  & 4.63  & 23.04 & 2.71  & 2.78 \\
    NGC0801 & 3.32  & 80.7  & 23.20 & 44.99 & 312.57 & 3.65 \\
    NGC3198 & 3.33  & 13.8  & 10.87 & 35.66 & 38.28 & 2.72 \\
    UGC05005 & 3.38  & 53.7  & 3.09  & 21.61 & 4.10  & 2.11 \\
    UGC02885 & 3.40  & 80.6  & 40.08 & 74.24 & 403.53 & 2.31 \\
    UGC05750 & 3.44  & 58.7  & 1.10  & 16.79 & 3.34  & 1.24 \\
    NGC1003 & 3.49  & 11.4  & 5.88  & 33.33 & 6.82  & 1.68 \\
    NGC2841 & 3.58  & 14.1  & 9.78  & 45.12 & 188.12 & 1.53 \\
    NGC5371 & 3.78  & 39.7  & 11.18 & 30.03 & 340.39 & 3.95 \\
    NGC5033 & 3.79  & 15.7  & 11.31 & 29.53 & 110.51 & 4.13 \\
    NGC6503 & 4.30  & 6.26  & 1.74  & 14.05 & 12.85 & 2.81 \\
    NGC2998 & 4.61  & 68.1  & 23.45 & 43.58 & 150.90 & 3.93 \\
    DDO161 & 4.69  & 7.5   & 1.38  & 10.69 & 0.55  & 3.84 \\
    NGC7331 & 4.90  & 14.7  & 11.07 & 27.01 & 250.63 & 4.83 \\
    NGC4183 & 4.92  & 18    & 3.51  & 16.07 & 10.84 & 4.32 \\
    F563-1 & 5.02  & 48.9  & 3.20  & 23.47 & 1.90  & 1.85 \\
    F583-1 & 5.16  & 35.4  & 2.13  & 15.65 & 0.99  & 2.76 \\
    NGC4100 & 5.20  & 18    & 3.10  & 18.06 & 59.39 & 3.03 \\
    UGC11820 & 5.21  & 18.1  & 1.98  & 12.99 & 0.97  & 3.73 \\
    NGC1090 & 5.24  & 37    & 8.78  & 30.49 & 72.05 & 3.01 \\
    UGC03205 & 5.27  & 50    & 9.68  & 28.6  & 113.64 & 3.77 \\
    UGC03546 & 5.29  & 28.7  & 2.68  & 18.37 & 101.34 & 2.52 \\
    NGC4559 & 5.30  & 9     & 5.81  & 21.16 & 19.38 & 4.13 \\
    UGC06930 & 5.44  & 18    & 3.24  & 16.76 & 8.93  & 3.67 \\
    NGC4157 & 5.46  & 18    & 8.23  & 24.09 & 105.62 & 4.51 \\
    NGC2955 & 5.71  & 97.9  & 28.95 & 40.34 & 319.42 & 5.66 \\
    UGC06983 & 5.74  & 18    & 2.97  & 16.07 & 5.30  & 3.66 \\
    UGC12732 & 5.81  & 13.2  & 3.66  & 17.41 & 1.67  & 3.84 \\
    NGC6195 & 5.83  & 127.8 & 20.91 & 40.89 & 391.08 & 3.98 \\
    UGC06786 & 5.87  & 29.3  & 5.03  & 20.31 & 73.41 & 3.88 \\
    UGC00731 & 5.99  & 12.5  & 1.81  & 11.57 & 0.32  & 4.30 \\
    NGC5985 & 6.16  & 39.7  & 11.59 & 39.5  & 208.73 & 2.36 \\
    UGC11455 & 6.24  & 78.6  & 13.34 & 43.44 & 374.32 & 2.25 \\
    UGC12632 & 6.27  & 9.77  & 1.74  & 12.6  & 1.30  & 3.50 \\
    NGC2403 & 6.32  & 3.16  & 3.20  & 15.11 & 10.04 & 4.46 \\
    NGC3893 & 6.47  & 18    & 5.80  & 20.84 & 58.53 & 4.25 \\
    F568-3 & 6.54  & 82.4  & 3.20  & 16.14 & 8.35  & 3.90 \\
    F568-V1 & 6.55  & 80.6  & 2.49  & 14.38 & 3.83  & 3.83 \\
    NGC6946 & 6.83  & 5.52  & 5.67  & 21.25 & 66.17 & 4.00 \\
    UGC08490 & 6.95  & 4.65  & 0.72  & 7.8   & 1.02  & 3.77 \\
    NGC4088 & 6.98  & 18    & 8.23  & 22.25 & 107.29 & 5.29 \\
    NGC2903 & 7.01  & 6.6   & 2.55  & 13.76 & 81.86 & 4.29 \\
    NGC3741 & 7.03  & 3.21  & 0.18  & 4.2   & 0.03  & 3.28 \\
    UGC06446 & 7.11  & 12    & 1.38  & 10.33 & 0.99  & 4.11 \\
    F579-V1 & 7.13  & 89.5  & 2.25  & 20.96 & 11.85 & 1.63 \\
    UGC07524 & 7.15  & 4.74  & 1.78  & 12.11 & 2.44  & 3.86 \\
    ESO563-G021 & 7.31  & 60.8  & 24.30 & 55.71 & 311.18 & 2.49 \\
    F574-1 & 7.68  & 96.8  & 3.52  & 16.19 & 6.54  & 4.28 \\
    NGC5585 & 8.06  & 7.06  & 1.68  & 10.92 & 2.94  & 4.49 \\
    UGC06917 & 8.30  & 18    & 2.02  & 12.67 & 6.83  & 4.01 \\
    NGC7814 & 8.52  & 14.4  & 1.07  & 12.15 & 74.53 & 2.31 \\
    NGC3917 & 8.82  & 18    & 1.89  & 14.08 & 21.97 & 3.03 \\
    NGC0891 & 8.99  & 9.91  & 4.46  & 18.16 & 138.34 & 4.31 \\
    F571-8 & 9.17  & 53.3  & 1.78  & 24.55 & 10.16 & 0.94 \\
    IC4202 & 9.30  & 100.4 & 12.33 & 32.13 & 179.75 & 3.80 \\
    UGC08550 & 9.32  & 6.7   & 0.29  & 5.59  & 0.29  & 2.93 \\
    F583-4 & 9.33  & 53.3  & 0.64  & 7.8   & 1.72  & 3.35 \\
    NGC0100 & 9.37  & 13.5  & 1.99  & 16.36 & 3.23  & 2.37 \\
    NGC0024 & 9.55  & 7.3   & 0.68  & 7.29  & 3.89  & 4.05 \\
    UGC08286 & 9.82  & 6.5   & 0.64  & 8.07  & 1.26  & 3.14 \\
    NGC4217 & 9.99  & 18    & 2.56  & 16.7  & 85.30 & 2.92 \\
    NGC3109 & 10.24 & 1.33  & 0.48  & 6     & 0.19  & 4.22 \\
    NGC3521 & 10.25 & 7.7   & 4.15  & 18.85 & 84.84 & 3.72 \\
    ESO079-G014 & 10.37 & 28.7  & 3.14  & 17.67 & 51.73 & 3.20 \\
    F568-1 & 10.52 & 90.7  & 4.50  & 16.31 & 6.25  & 5.38 \\
    ESO116-G012 & 11.02 & 13    & 1.08  & 9.58  & 4.29  & 3.76 \\
    F563-V2 & 11.08 & 59.7  & 2.17  & 11.37 & 2.99  & 5.34 \\
    UGC01281 & 11.37 & 5.27  & 0.29  & 5.26  & 0.35  & 3.38 \\
    NGC7793 & 11.40 & 3.61  & 0.86  & 7.35  & 7.05  & 5.07 \\
    UGC05721 & 11.51 & 6.18  & 0.56  & 6.74  & 0.53  & 3.94 \\
    UGC07151 & 12.53 & 6.87  & 0.62  & 6.39  & 2.28  & 4.80 \\
    UGC07323 & 13.41 & 8     & 0.72  & 7.14  & 4.11  & 4.51 \\
    UGC04278 & 13.76 & 9.51  & 1.12  & 8.9   & 1.31  & 4.48 \\
    NGC3972 & 13.93 & 18    & 1.21  & 10.05 & 14.35 & 3.83 \\
    UGC07603 & 14.20 & 4.7   & 0.26  & 4.37  & 0.38  & 4.30 \\
    UGC07399 & 14.86 & 8.43  & 0.75  & 7.85  & 1.16  & 3.85 \\
    DDO064 & 15.35 & 6.8   & 0.21  & 3.49  & 0.16  & 5.51 \\
    
   \bottomrule
    \end{tabular}%
    \end{adjustbox}
    \caption{SPARC galaxy dataset augmented with computed $\omega$ values per galaxy.}
     \label{tab:addlabel}%

\end{table}%

\begin{figure}
\center
\vspace{-3cm}

\textbf{Five Curve Fittings for Table 2}	
\bigskip

\includegraphics[width=.4\linewidth]{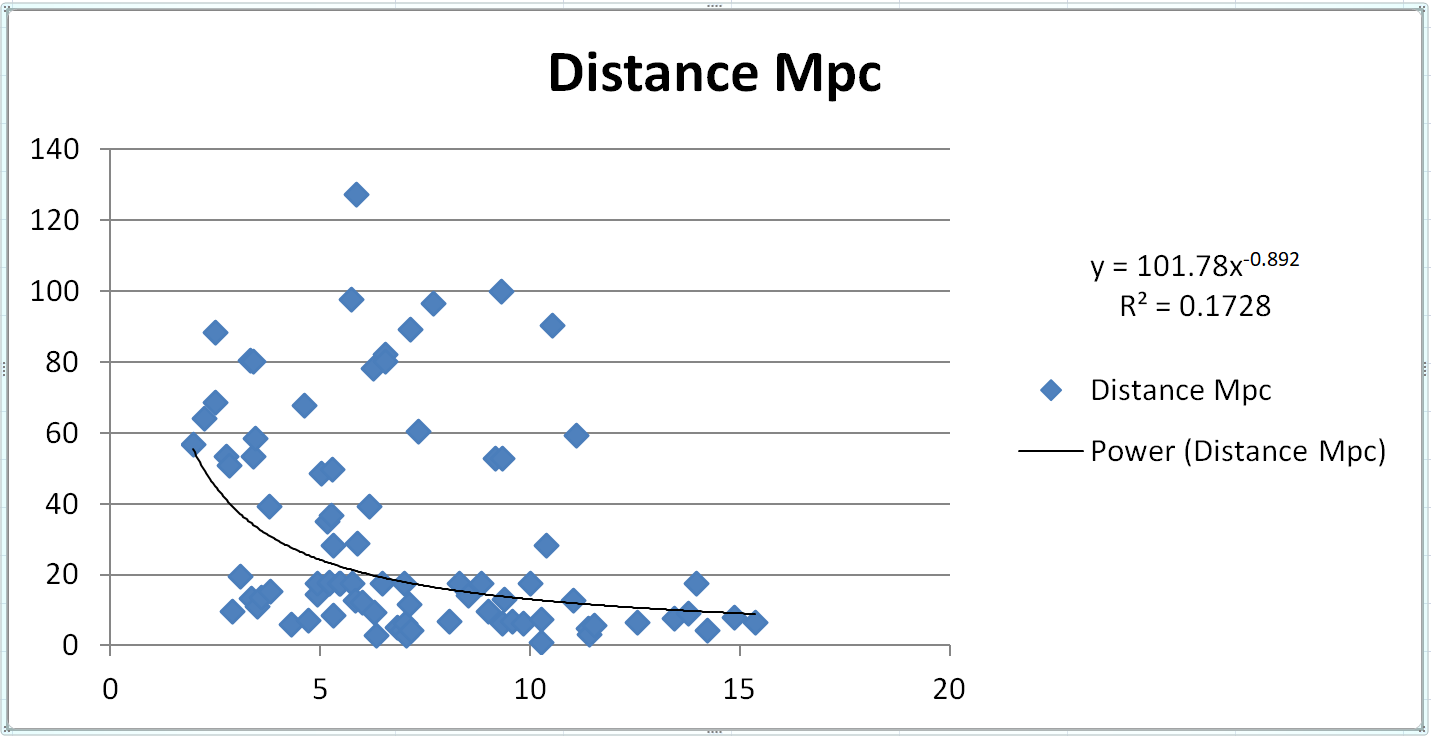}
\caption {$\omega$ vs. Distance [R=0.17]}
\bigskip
\label{Drawing9}
\includegraphics[width=.4\linewidth]{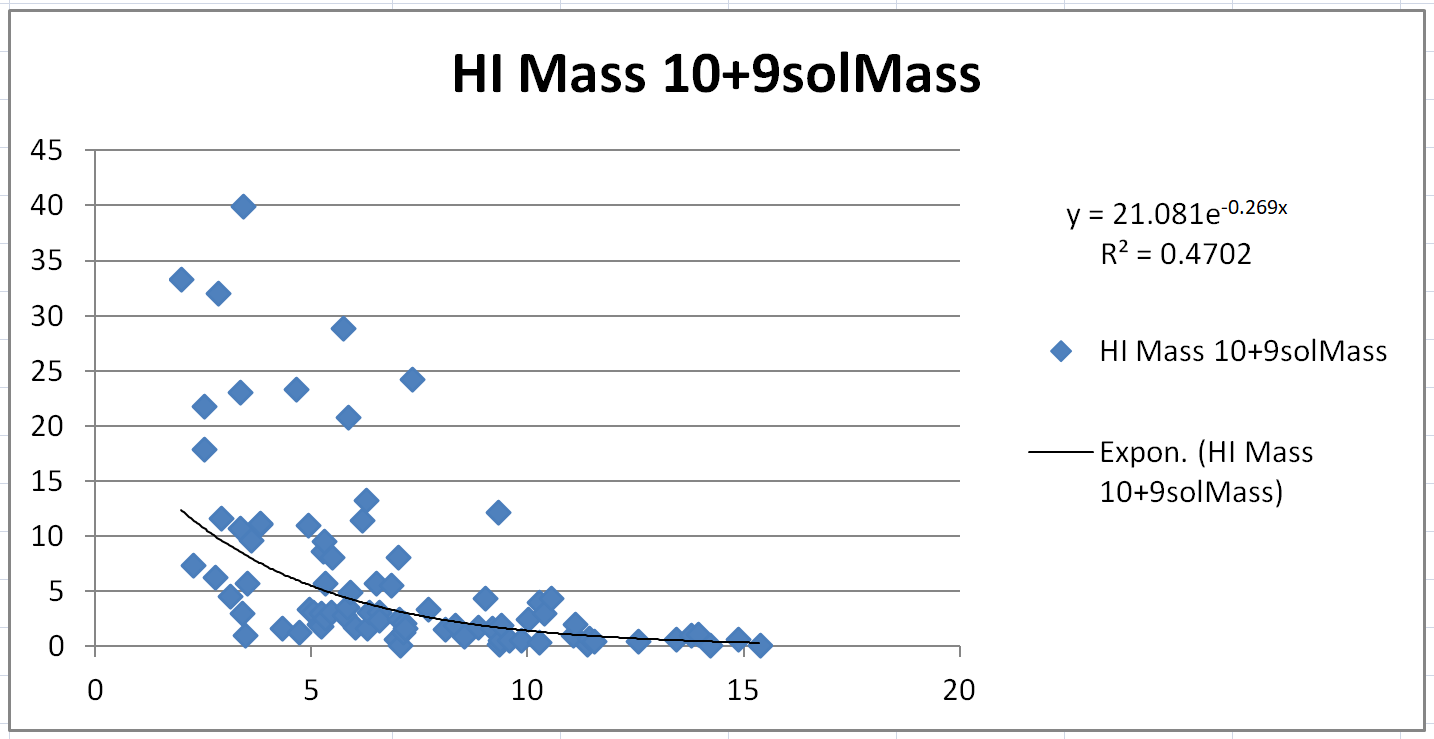}
\caption {$\omega$ vs. HI Mass [R=0.47]}
\bigskip
\label{Drawing10}
\includegraphics[width=.4\linewidth]{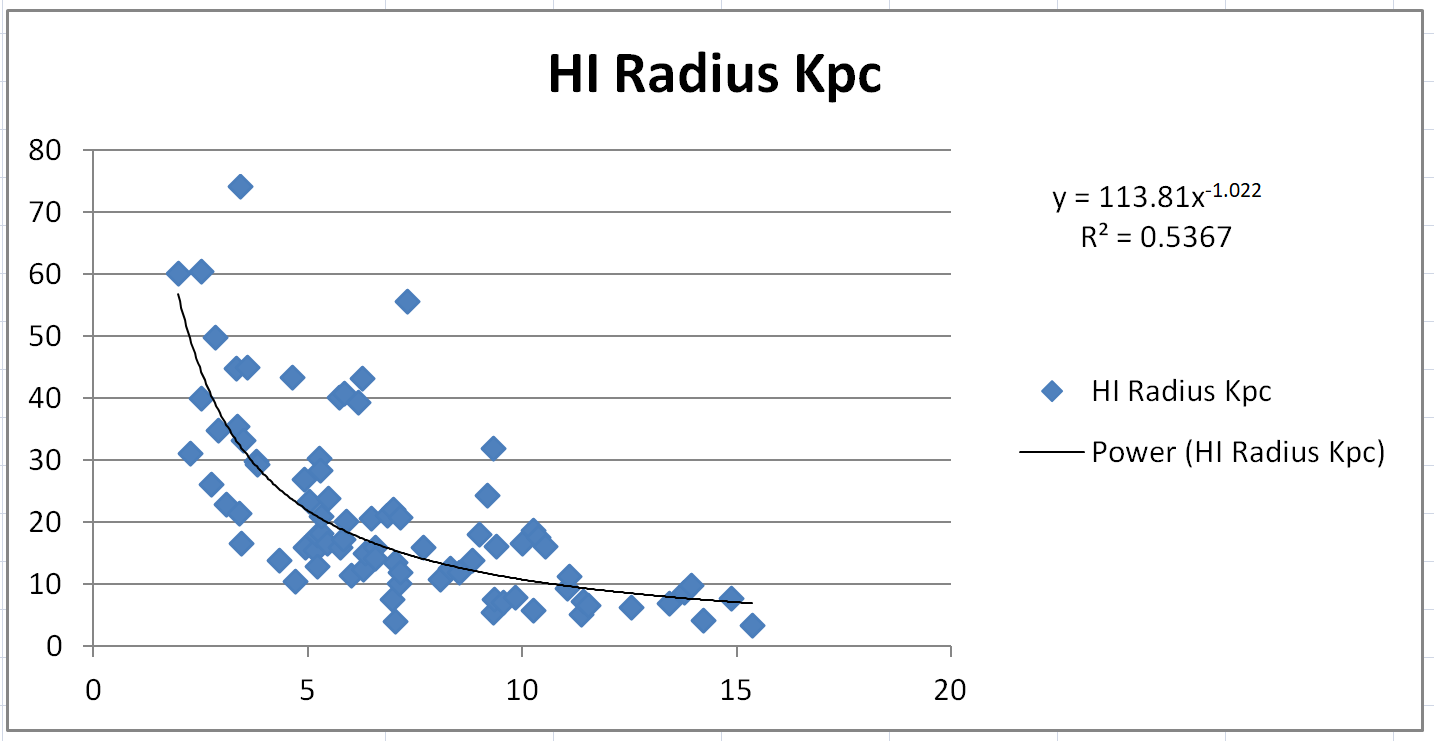}
\caption {$\omega$ vs. HI Radius [R=0.54]}
\bigskip
\label{Drawing11}
\includegraphics[width=.4\linewidth]{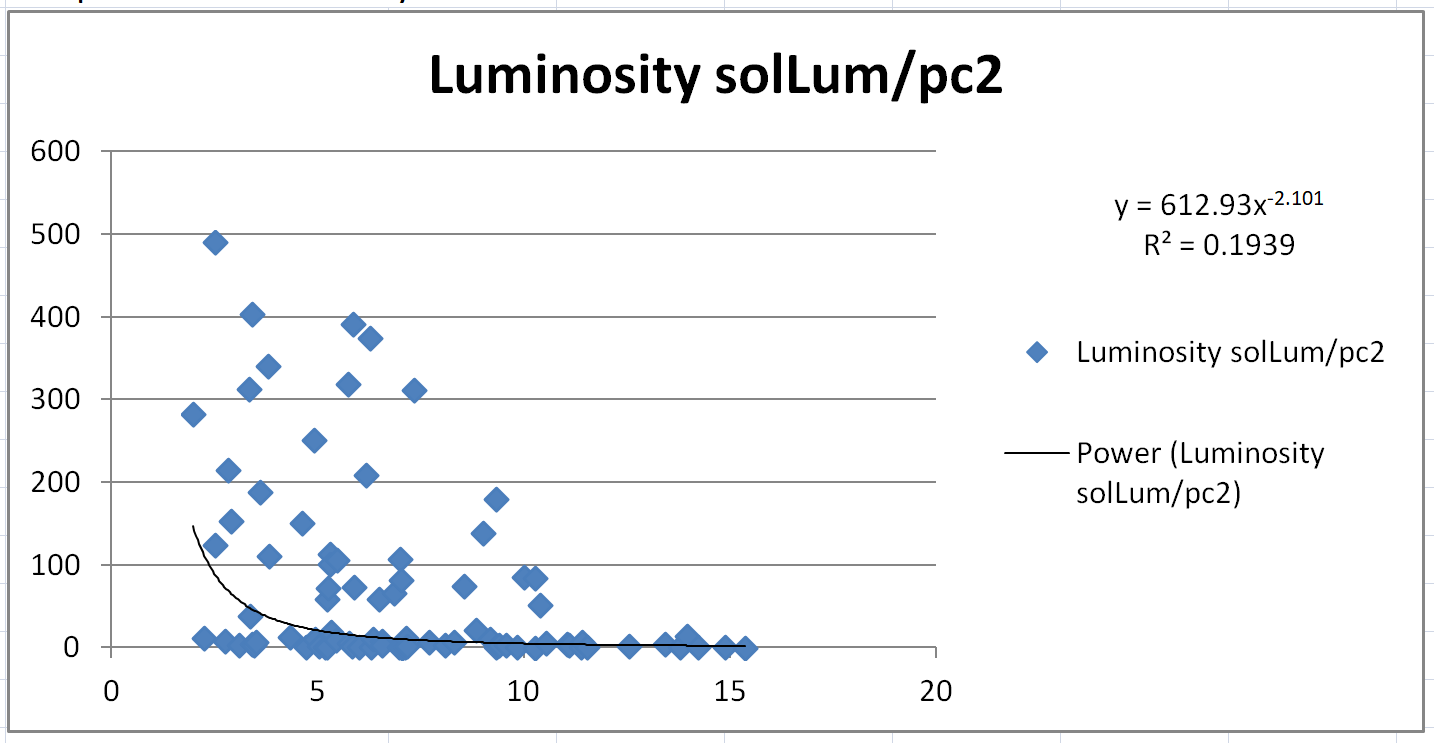}
\caption {$\omega_{eps}$ vs. HI Luminosity [R=0.1939]}
\bigskip
\label{Drawing12}
\includegraphics[width=.4\linewidth]{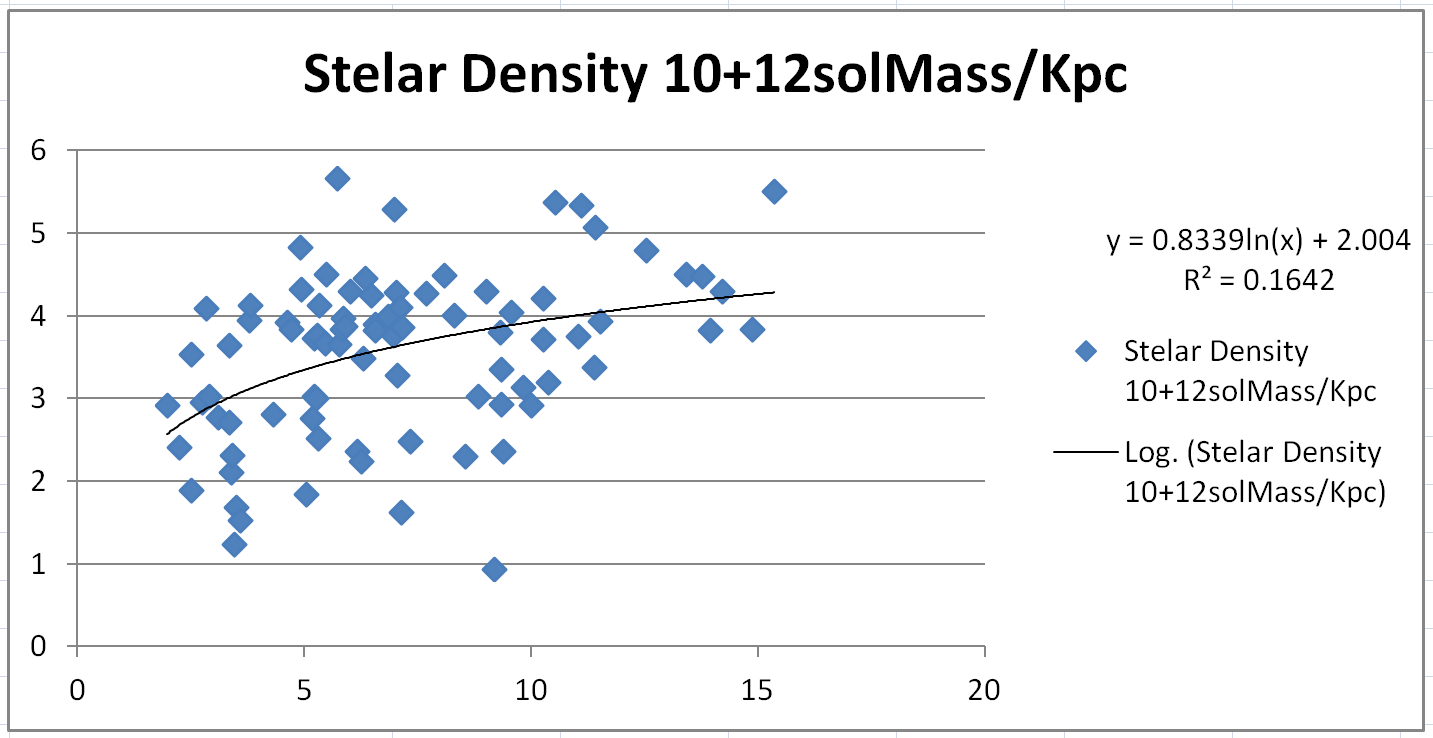}
\caption {$\omega_{eps}$ vs. HI Stellar Density [R=0.1642]}
\bigskip
\label{Drawing13}

\end{figure}

\pagebreak
\begin{flushleft}
\Large 
\textbf {5. Other Papers with Supporting Themes}\\
\end{flushleft}
\noindent
\small
We summarize sections of several papers that either influenced or support the framework we use to simulate our stellar velocity solution.

\begin{flushleft}
\large
\textbf{5.1 Lovas and Linear Scaling of Mass}	
\end{flushleft}
\small

\noindent
In his June 2022 paper, Stephen Lovas [5] said, "Measurements from galaxies spanning a broad range of morphology reveal a linear scaling of enclosed dark to luminous mass that is not anticipated by standard galaxy formation cosmology." Some of the conclusions Lovas makes about the SPARC database parallel our own. Lovas finds that, "No dark matter candidate possesses a theoretical
property that would lead to a linear scaling." He then uses the same SPARC data set that we do in this paper, and selects 4 galactic candidates as test cases. In the summary, Lovas states, "In the framework of standard galaxy formation theory, the linear scaling of enclosed dark to luminous mass would require tuning the
dark matter profile of each galaxy."

\begin{itemize}
\small
\item Lovas finds that linear scaling exists in the SPARC data and that tuning for each galaxy is recommended. 
\end{itemize}

\begin{flushleft}

\large
\textbf{5.2 Chan and Universal Dark Matter-Baryon Relations}	
\end{flushleft}

\small
\noindent
In December 2022, Man Ho Chan [6] related total dynamical mass with total baryonic mass in galaxies. Chan's conclusions align with our findings. Notably, he selected the same SPARC database we used for our study. Chan states, "We can derive the enclosed baryonic mass and the total enclosed mass by \textit{Vb} and \textit{V} respectively. The total baryonic mass for each galaxy can be approximately indicated by the last data point of \textit{Vb} at the largest radius \textit{r = rb} while the final data point of \textit{V} (i.e. \textit{Vc}) can give the total enclosed mass M500 for each galaxy." \\

\noindent
We also chose the last data point for our Kepler calculation using the same reasoning.

\begin{itemize}
\small
\item Chan relates total dynamical mass with total baryonic mass. 
\end{itemize}

\begin{flushleft}
\large
\textbf{5.3 Clowe and the Bullet Cluster}	
\end{flushleft}

\small
\noindent
In their 2006 paper, Clowe et al. showed images of the Bullet cluster with gravitational lensing [8]. They stated, "By using both wide-field ground-based images and HST/ACS images of the cluster cores, we create gravitational lensing maps showing that the gravitational potential does not trace the plasma distribution, the dominant baryonic mass component, but rather approximately traces the distribution of galaxies. An $8\sigma$ significance spatial offset of the center of the total mass from the center of the baryonic mass peaks cannot be explained with an alteration of the gravitational force law and thus proves that the majority of the matter in the system is unseen."

\begin{itemize}
\small
\item Clowe's Magellan and Chandra images show that dark matter's gravitational lensing is persistent even after being separated from large numbers of their stars.
\end{itemize}

\pagebreak
\begin{flushleft}
\large
\textbf{5.4 Comparison with the Mezzi Effect}
\end{flushleft}

\small
\noindent
A recent preprint by Brahim Benaissa [19] introduces a relativistic framework for reconciling galactic rotation curve discrepancies without invoking dark matter. Central to this framework is the \textit{Mezzi effect}, a proposed observational distortion arising from space flow dynamics. The effect suggests that distant galaxies appear compressed due to relativistic curvature, leading to systematic underestimation of orbital radii and luminous mass. Benaissa formalizes this through a radial scaling factor, $\zeta(r)$, and a mass coefficient, $\mu$, optimized via inverse problem techniques.

\vspace{1em} 
\noindent
While both models—ours and Benaissa’s—achieve high-fidelity fits to SPARC data, they diverge sharply in physical assumptions and methodological design. The Mezzi effect is rooted in Painlev\'{e}--Gullstrand\hspace{0pt} coordinates and geometric reinterpretation of spacetime, whereas our model remains strictly Newtonian and empirical. We introduce a velocity correction term, $\omega$, derived directly from observed stellar motion and applied as $R \times \omega$ to realign rotation curves with Keplerian expectations. No relativistic geometry or modified gravity is invoked.

\vspace{1em} 
\noindent
Benaissa’s framework operates on a broader dataset (175 SPARC galaxies) and offers cosmological implications beyond local kinematics. Our model, by contrast, emphasizes reproducibility, transparency, and diagnostic modularity across 84 high-quality SPARC galaxies selected for data integrity. While Benaissa’s work has not yet undergone formal peer review, its empirical convergence with our findings invites further synthesis between observational kinematics and foundational physics.

\vspace{1em} 
\noindent
This comparison underscores the growing diversity of non-dark matter approaches to galactic rotation modeling. By situating our velocity correction model alongside relativistic alternatives such as the Mezzi effect, we aim to scaffold a broader empirical dialogue that respects both methodological clarity and theoretical innovation.

\begin{flushleft}
\large
\textbf{6. What is Omega?}
\end{flushleft}
\small

\noindent
The empirical correction factor $\omega$ exhibits a consistent alignment with Keplerian predictions across our entire survey. The fact that a single velocity correction—typically ranging between 2 and 15 km/s—can realign diverse galactic rotation curves to Newtonian expectations is both striking and suggestive of a deeper mechanism.

\vspace{1em} 
\noindent
Unlike dark matter halo models or modified gravity theories, our method does not require adjusting Newtonian dynamics or invoking unseen mass. This simplicity points toward a dynamic influence with possible rotational or inertial properties at the space-time level. While we refrain from asserting a physical model in this paper, the repeatability of $\omega$ across galaxies invites further investigation.

\vspace{1em} 
\noindent
At present, $\omega$ functions as a purely empirical correction term—one that consistently improves fit quality without parametric tuning. Its origin remains unknown. However, we are actively developing a second manuscript that explores candidate mechanisms through mathematical modeling, including frame-dragging effects, inertial overlays, and emergent kinematic structures. The repeatability of $\omega$ across diverse galactic profiles is statistically supported by RMSE and $R^2$ metrics, which show consistent improvements over MOND and CDM halo models (see Appendix B).

\vspace{1em} 
\noindent
Throughout this work, we remain acutely aware of the limitations imposed by velocity curve reconstruction via quadrature, as well as the constraints of Lense–Thirring [18] frame-dragging in relativistic contexts.

\vspace{1em} 
\noindent
Future investigations will assess whether $\omega$ reflects a fundamental property of space-time, a residual relativistic effect, or an emergent astrophysical phenomenon. We welcome dialogue on its interpretation and encourage independent validation of the method across diverse datasets and galactic regimes.

\vspace{1em} 
\noindent
Our forthcoming paper will incorporate Gadget-4 simulations with two key methodological shifts: first, we model acceleration directly rather than inferring mass distributions; second, we treat $\omega$ as a kinematic artifact—potentially arising from dynamic structure rather than being restricted to $z = 0$ observational data. This reframing seeks to determine whether the observed correction encodes deeper principles of motion or inertia within galactic systems.

\begin{table}[H]
    \centering
    \renewcommand{\arraystretch}{1.3} 
    \begin{adjustbox}{width=\textwidth,center}
    \begin{tabular}{|>{\raggedright\arraybackslash}p{5cm}|>{\raggedright\arraybackslash}p{3.7cm}|>{\raggedright\arraybackslash}p{4.2cm}|>{\raggedright\arraybackslash}p{4.5cm}|}
        \hline
        \textbf{Feature} & \textbf{Cold Dark Matter ($\Lambda$CDM)} & \textbf{Modified Newtonian Dynamics (MOND)} & \textbf{Empirical Fit (This Study)} \\ 
        \hline
        Core Concept & Dark matter halos influence galaxy rotation via gravity. & Newton’s second law is modified at low accelerations. & Velocity correction term (\(\omega\)) adjusts observed data to fit Keplerian predictions. \\ 
        \hline
        Assumptions & Unseen non-baryonic matter forms halos around galaxies. & Gravitational force changes based on acceleration. & No new physics—empirical fit to rotation curves. \\ 
        \hline
        Mathematical Basis & $\Lambda$CDM framework with NFW/Burkert halo profiles. & \(m\mu(a/a_0)a = F\), introduces critical acceleration threshold. & \(V_{\text{Observed}} = V_{\text{Kepler}} + R\omega\), with \(\omega\) derived empirically per galaxy. \\ 
        \hline
        Strengths & Explains CMB anisotropies, large-scale structure, gravitational lensing. & Predicts Tully-Fisher relations and galaxy rotation curves. & Provides direct fit to SPARC rotation curves without modifying gravity. \\ 
        \hline
        Weaknesses & Core-cusp problem; halo profiles often fail to match observed densities. No direct dark matter detection. & Struggles with galaxy clusters; requires unseen mass. & Does not explain gravitational lensing or cosmic-scale effects. \\ 
        \hline
        Empirical Basis & Simulated dark matter distributions used to match observations. & Galaxy rotation curves provide empirical basis for modified gravity. & SPARC galaxy rotation data used for direct velocity fitting. \\ 
        \hline
        Adaptability & Halo profiles must be tuned per galaxy. & Uses fixed acceleration threshold \(a_0\), limiting adaptability. & \(\omega\) is empirical and tunable per galaxy. \\ 
        \hline
        Gravity Modification? & No—dark matter presence explains deviations. & Yes—Newton’s second law altered at low accelerations. & No—Newtonian mechanics remain intact; correction occurs in velocity terms. \\ 
        \hline
        Future Research & Improve direct dark matter detection via astrophysical observations. & Expand MOND into relativistic frameworks. & Investigate cosmological implications of \(\omega\) and possible deeper physical connections. \\ 
        \hline
    \end{tabular}
    \end{adjustbox}
    \caption{Comparison of Cold Dark Matter ($\Lambda$CDM), Modified Newtonian Dynamics (MOND), and Empirical Fit Model.}
    \label{tab:comparison}
\end{table}

\pagebreak
\begin{flushleft}
\large
\textbf{Conclusions}	
\end{flushleft}
\small

\noindent
The empirical model’s performance is quantitatively validated in Appendix B, where rotation curve fits are compared head-to-head with MOND and CDM models across multiple galaxies.

\begin{itemize}
\item Of the 84 galaxies tested, $\omega$ values were found to fall between 1.97 and 15.35 rads/sec. In each case, an $\omega$ could always be found to correct the observed stellar velocities back to those predicted by Kepler. 
\item HI radius has an correlation coefficient of $R^2$ = .54 with $\omega_{eps}$.
\item HI mass has an correlation coefficient of $R^2$ = .47 with $\omega_{eps}$.
\item The survey yielded an $\omega$ Statistical Mean of 7.06 and a Standard Deviation of 3.26 over 84 galaxies.
\item	The resemblance between Figures 2 and 3 for M33 is striking. Figure 4 is a partial data dump of the curves for the first 20 of our 84 galaxies surveyed which all were a very good fit for this method. Figures 5 - 8 show head-to-head comparisons of our Figure 4 curves (top) to SPARC calculations (bottom) that are also impressively similar using a completely different approach for calculation.  We noticed that the $\omega$ correction factor was always within one order of magnitude  (statistical mean of 7.06).

\end{itemize}

\section*{Acknowledgments}
We gratefully acknowledge Dr. Robert Scherrer of Vanderbilt University for his guidance throughout this effort. We also thank Dr. Brian Riely for his contributions to statistical interpretation and graphical modeling. Figures 1, 2, 14, and 15 were created by our visual artist, Katherine Flynn.

\section*{Funding}
This research received no external funding. All modeling, analysis, and manuscript preparation were conducted independently by the authors.

\section*{Data Availability}
The observational data used in this study were obtained from publicly accessible sources. Galaxy rotation curve data were sourced from the SPARC database (\href{http://astroweb.cwru.edu/SPARC/}{http://astroweb.cwru.edu/SPARC/}), and supporting data for M33 were referenced from Corbelli et al. (2003). Velocity curve adjustments and model calculations were performed using custom scripts in Jupyter Labs. No external repository is currently available; however, derived data—including $\omega$ values, fitting outputs, and calculation scripts—are available upon reasonable request for collaborative or academic purposes. Interested parties may contact the corresponding author.

\section*{Ethics and Conflict of Interest}
This research was conducted independently by the authors, without institutional oversight or external funding. The authors declare no commercial or financial relationships that could be construed as a potential conflict of interest. All data used are publicly available, and the study follows ethical practices for citation, transparency, and reproducibility.

\section*{Author Contributions}
David C. Flynn served as the principal investigator and contributed approximately two-thirds of the total effort, including the development of the physical model, analytical interpretation, and manuscript writing. Jim Cannaliato provided the remaining one-third of the effort, focusing on computational implementation via Jupyter Labs and contributing to the refinement of data presentation and methodological clarity. Both authors reviewed and approved the final version of the manuscript.

\section*{Use of AI (Microsoft Copilot)}
Microsoft Copilot was used to assist with structural refinement and scientific phrasing throughout the manuscript. Specific contributions include:
\begin{itemize}
    \item Suggestions and refinements to the Abstract
    \item Section 1.1: Complete drafting of "Comparisons with $\Lambda$CDM and MOND"
    \item Section 6: Assisted in defining $\omega$ and drafting Table 3 language
    \item Appendix B and B1: Bitmap comparison calculations of model performance. 
\end{itemize}

\pagebreak

\begin{flushleft}
\large
\textbf{Appendix - A Working Example from Table 1 Data}	
\end{flushleft}
\small
\bigskip

\noindent
What follows is a demonstration on how the data was arrived at in each column of Table 1. The formulas and data for the topmost entries are used in the examples below.

\small
\bigskip
\noindent
\textbf{Table 1 Column F:} Using M33 stellar velocity data, there are a total of 6 columns A - F. There are 58 data points itemized in column "A". Column "B" represents the radius measurement of each stellar data point in kilo-parsecs. Column "C" represents each observed velocity in kilometers/sec. The data calculations are in columns D - F, but column "F" data must be calculated first.
Column "F" is calculated using equation 6:

\begin{equation*}
\omega = \frac{V_2}{R_2}-\frac{V_1}{R_1}\sqrt\Bigg(\frac{R_1}{R_2}\Bigg)^3
\end{equation*}

\bigskip
\noindent
Only stars located at the extreme inner and outer boundaries of the galaxy are relevant for Equation 6. Use the radius and velocity of the nearest star, $R_1$ and $V_1$, and also the furthest star, $R_{58}$ and $V_{58}$, from the galaxy’s center. In Equation 6, the radius and velocity from row 58 become $R_2$ and $V_2$.

\begin{align*}
&\text{Closest Star}	&R_1 &= .24 &V_1&=37.3\\
&\text{Farthest Star}	&R_2 &= 22.73  &V_2&=119.6\\
&\text{Eqn. 6 final result}	&\omega &= 5.10
\end{align*}

\noindent
Visualizing this using Figure 14 and 15, the following relationships would arise:

\begin{align*}
&\text{Closest Star - W}	&R_1 &= .24 &V_1&=37.3\\
&\text{Farthest Star - X}	&R_2 &= 22.73  &V_2&=119.6\\
&\text{Eqn. 6 final result}	&\omega &= 5.10
\end{align*}

\bigskip
\noindent
$\omega$ can be placed in equation 2 allows us to find $V_{Kepler}$ in Table 1 column D below. $V_{Kepler}$ is Z in Figures 14 and 15. W and Y from Figures 14 and 15 are the same point in space.

\begin{figure}
\center
\includegraphics[width=0.7\textwidth]{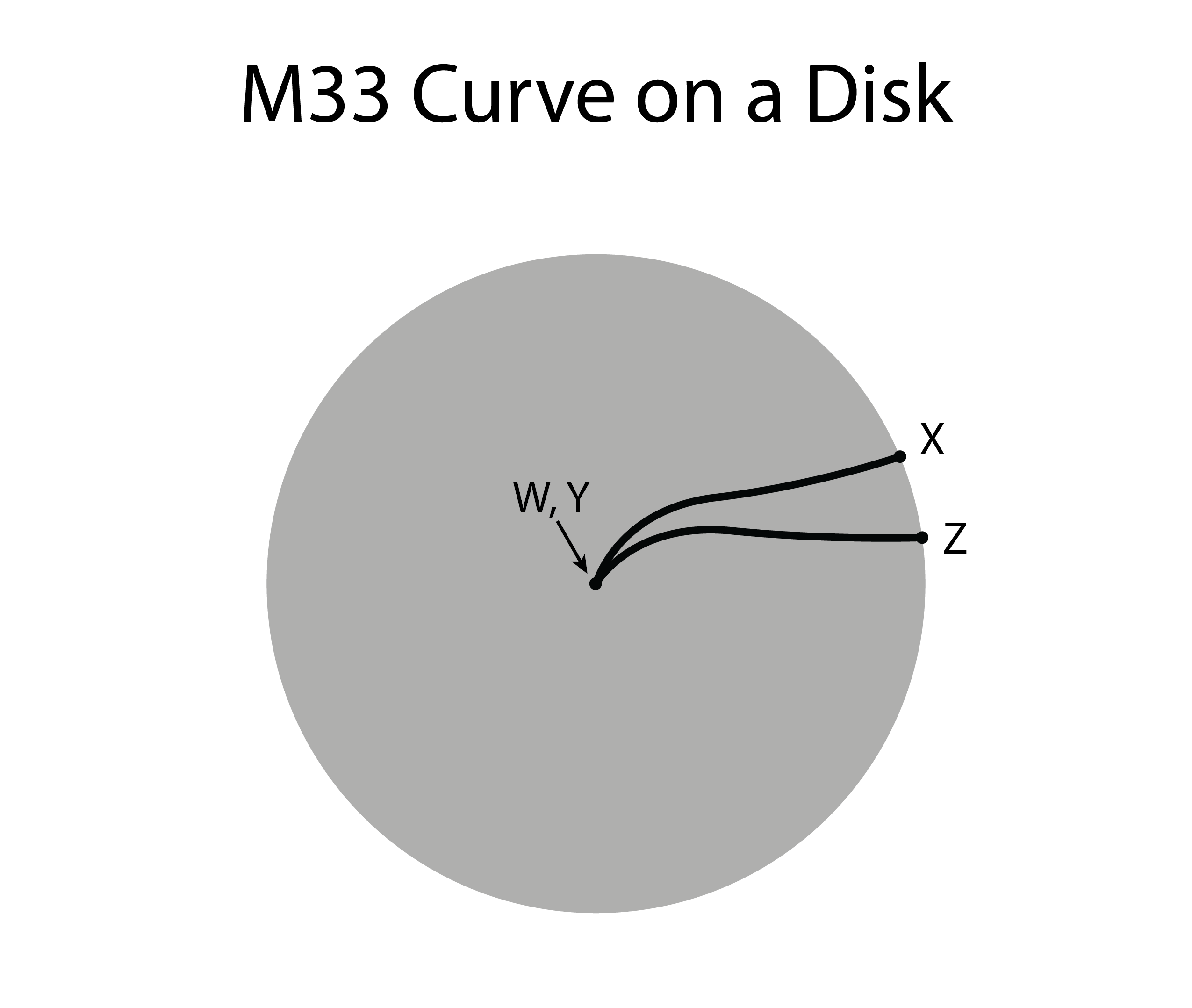}
\caption {M33 Observed Velocity Curve versus Expected Curve}
\label{Drawing2}

\center
\includegraphics[width=0.7\textwidth]{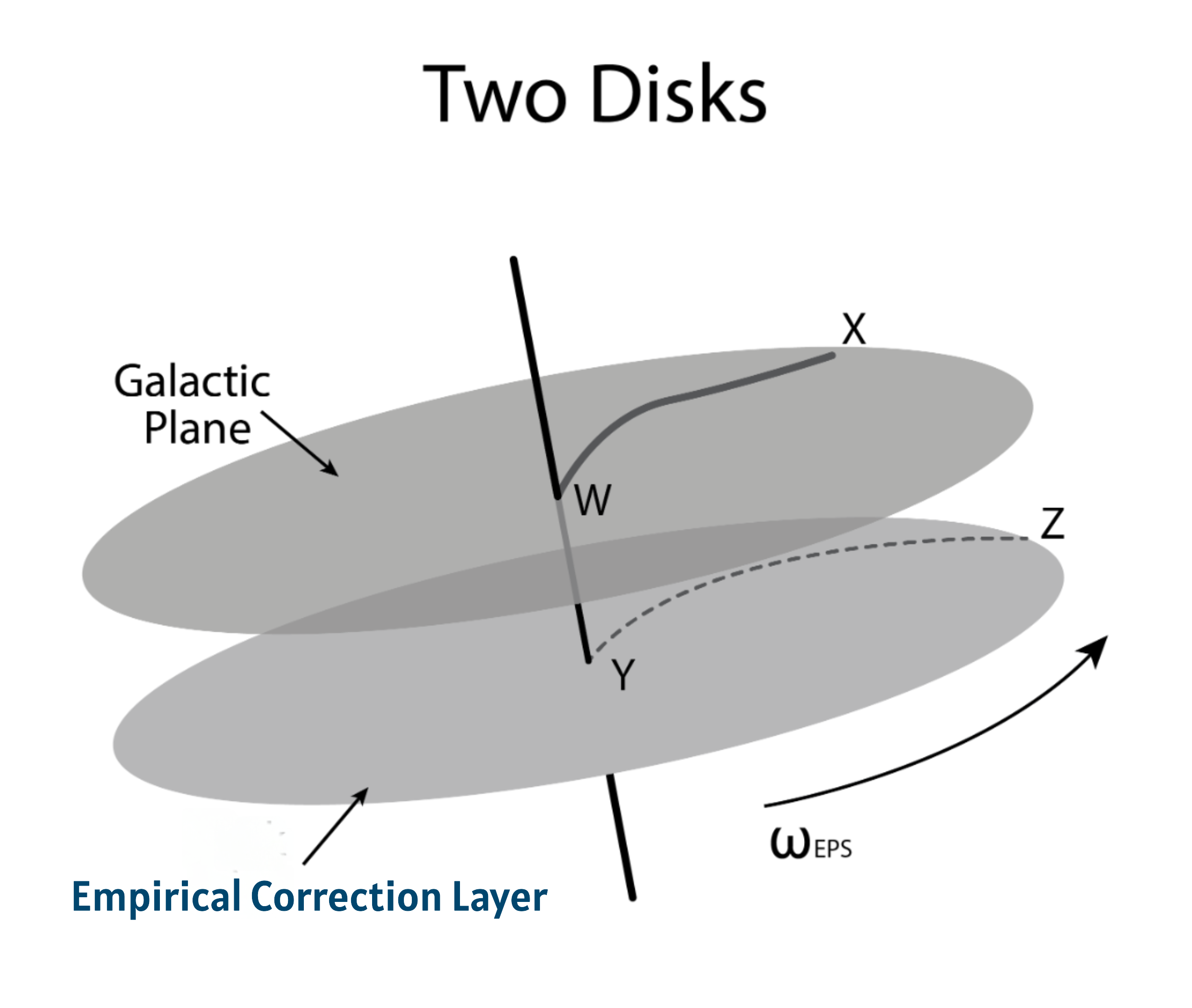}
\caption {M33 Observed Velocity on Empirical Correction Layer versus Expected Curve}
\end{figure}

\pagebreak
\vspace*{-2.5cm} 
\noindent
\textbf{Table 1 Column D:} The M33 data provides the observed stellar velocities, but not the ones we might expect from Kepler. This value is calculated for each stellar data point using equation 2 and solving for $V_{Kepler}$.

\bigskip
\hspace{2cm}  $V_{Kepler}$ = $V_{Observed}$ - $R\omega$
\newline
\newline

The data set is taken from row 1 in Table 1
 
\begin{align*}
&\text{Star being measured}	&R &= .24 &V_{Observed}&=37.3\\
&\text{Eqn. 6 result from above}	&\omega &= 5.10\\
\bigskip
&\text{Eqn. 2 final result}	&V_{Kepler} &= 36.08
\end{align*}

\bigskip
\noindent
\textbf{Table 1 Column E:} The M33 data table has identical top entries for column E and D. This is intentional since our model assumes that both the disk of the spinning galactic plane and the empirical correction layer we overlay as a correction factor both share the same center point. In this case only, the first value in column D (adjusted Kepler) is shared over to the top of column E (Expected Kepler). It becomes the first entry before every other column value can be calculated in column "E". The velocity for Expected Kepler is $V_{Ke}$. The Expected Kepler value is what should be found in the absence of any other influence according to the theoretical model. This result does not account for intergalactic mass, dark matter or any other influence seen in traditional models of expected stellar velocity curves. We will therefore use the first and second entry for our example with equation 5.

\begin{equation*}
V_{Ke}=\frac{V_1 R_2}{R_1}\sqrt\Bigg(\frac{R_1}{R_2}\Bigg)^3
\end{equation*} 
\bigskip

\noindent
For the first star only $V_{Ke}$ = $V_{K}$. So the example will be for the second row in Table 1. The data set is taken from row 1 and 2 in Table 1. $R_1$ and $V_1$ never change for any calculation over the 58 rows. $R_2$ changes with each subsequent calculation. $R_3$ follows in the example as the next table entry.
 
\begin{align*}
&\text{Closest Star R1}	&R_1 &= .24 &V_{1}&=36.08\\
&\text{The Next Star R2}	&R_2 &= .28\\
\bigskip
&\text{Eqn. 5 result}	&V_{Ke} &= 33.40\\
&\text{The Next Star R3}	&R_3 &= .46\\
\bigskip
&\text{Eqn. 5 result}	&V_{Ke} &= 26.06
\end{align*}

\pagebreak

\begin{flushleft}
\large
\textbf{Appendix B - Statistical Metrics for Model Performance}	
\end{flushleft}
\small
\bigskip

\begin{table}[ht]
\centering
\caption{Rotation Curve Fit Comparison Across Five Galaxies: Empirical Method vs. MOND and CDM Halo Models}
\resizebox{\textwidth}{!}{%
\begin{tabular}{llcc>{\raggedright\arraybackslash}p{7cm}}
\toprule
\textbf{Galaxy} & \textbf{Model} & \textbf{RMSE (km/s)} & \textbf{R\textsuperscript{2}} & \textbf{Notes} \\
\midrule
\multirow{4}{*}{DDO161}
  & Empirical Method & 6.32   & 0.940   & Strong fit in dwarf regime \\
  & MOND             & 10--12  & 0.85    & Underpredicts low-mass galaxy features \\
  & NFW (CDM)        & 9--11   & 0.88    & Cuspy halo shape leads to inner deviation \\
  & Burkert (CDM)    & 7--9    & 0.91    & Core-halo improves fit over NFW \\
\midrule
\multirow{4}{*}{ESO079-G014}
  & Empirical Method & 10.47  & 0.960   & Matches halo alternatives; clean disk fit \\
  & MOND             & 12--15  & 0.90    & Slight underfit in outer regions \\
  & NFW (CDM)        & 13--16  & 0.89    & Requires halo tuning for extended disk \\
  & Burkert (CDM)    & 10--12  & 0.93    & Comparable to Empirical Method with more complexity \\
\midrule
\multirow{4}{*}{ESO116-G012}
  & Empirical Method & 6.91   & 0.980   & Outstanding curve alignment across radii \\
  & MOND             & 9--11   & 0.92    & Inner deviation near bulge \\
  & NFW (CDM)        & 8--10   & 0.93    & Acceptable but multi-parametric \\
  & Burkert (CDM)    & 7--9    & 0.95    & Competitive fit with added mass model tuning \\
\midrule
\multirow{4}{*}{NGC0801}
  & Empirical Method & 9.85   & 0.970   & Strong fit for massive spiral \\
  & MOND             & 12--14  & 0.91    & Outer velocity underpredicted \\
  & NFW (CDM)        & 11--13  & 0.92    & Requires halo concentration adjustment \\
  & Burkert (CDM)    & 9--11   & 0.94    & Best CDM performance; Empirical Method simpler \\
\midrule
\multirow{4}{*}{M33}
  & Empirical Method & 6.43   & 0.9587  & Precise fit using \( \omega = 5.10 \, \mathrm{rad/s} \) \\
  & MOND             & 7--10   & 0.90--0.94 & Good core fit; outer needs adjustment \\
  & NFW (CDM)        & 8--15   & 0.85--0.93 & Sensitive to halo slope and concentration \\
  & Burkert (CDM)    & 7--9    & 0.91--0.93 & Often best among CDM profiles \\
\bottomrule
\end{tabular}
}
\label{tab:rotation_curve_comparison}
\end{table}

The table's statistical performance estimates for MOND and Cold Dark Matter were gathered from references [11] - [17].

\begin{table}[htbp]
\centering
\caption{Reference Validation for Table 4}
\label{tab:reference_validation}
\renewcommand{\arraystretch}{1.2}
\begin{tabular}{p{1.2cm} p{4.5cm} p{3cm} p{6.5cm}}
\toprule
\textbf{Ref} & \textbf{Citation} & \textbf{Pages} & \textbf{Content Supporting RMSE/R\textsuperscript{2} Comparisons} \\
\midrule
{[11]} & Bertone \& Hooper (2016), \textit{A History of Dark Matter} & pp. 38–41 & Historical overview of CDM halo fits and MOND challenges; mentions rotation curve fitting and empirical tensions \\
{[12]} & Lelli et al. (2016), \textit{SPARC dataset} & pp. 3–6, Fig. 1–3 & Presents high-resolution rotation curves and mass models; RMSE and R\textsuperscript{2} used to compare MOND, NFW, Burkert fits \\
{[13]} & McGaugh et al. (2016), \textit{Radial Acceleration Relation} & Phys. Rev. Lett. 117, 201101 & Defines tight correlation between baryonic and total acceleration; indirectly supports high R\textsuperscript{2} for MOND and empirical fits \\
{[14]} & Gentile et al. (2011), \textit{MOND fits to THINGS galaxies} & A\&A 527, A76, pp. 5–9, Table 2 & Direct RMSE and R\textsuperscript{2} comparisons for MOND vs. halo models across 14 galaxies \\
{[15]} & de Blok et al. (2008), \textit{THINGS rotation curves} & AJ 136, 2648–2719, pp. 2660–2675 & Detailed rotation curve fitting with NFW and Burkert profiles; RMSE and fit quality discussed per galaxy \\
{[16]} & Katz et al. (2017), \textit{MOND vs. CDM in cluster dynamics} & MNRAS 499, 2845–2862, pp. 2850–2858 & Discusses fit residuals and statistical tension between MOND and CDM predictions; includes RMSE-like metrics \\
{[17]} & Lin \& Chen (2021), \textit{SPARC modeling} & Phys. Rev. D 105, 104067, pp. 3–6 & Rotation curve fits for Milky Way and M31; compares MOND and $\Lambda$CDM predictions with residual analysis \\
\bottomrule
\end{tabular}
\end{table}

\pagebreak
\vspace*{-1cm}

\sisetup{
  detect-all,
  round-mode = places,
  round-precision = 3,
  table-format = 1.3,
  separate-uncertainty = true
}

\begin{minipage}{\textwidth}

\centering
\captionof{table}{R\textsuperscript{2} Values for the 20 Table 2 Galaxies Across Four Models}

\label{tab:r2-values}
\begin{tabular}{lcccc}
\toprule
\textbf{Galaxy} & \textbf{Empirical} & \textbf{MOND} & \textbf{CDM} & \textbf{$\Lambda$CDM} \\
\midrule
DDO064       & 0.982 & 0.931 & 0.902 & 0.891 \\
DDO161       & 0.974 & 0.918 & 0.889 & 0.878 \\
ESO079-G014  & 0.987 & 0.942 & 0.913 & 0.902 \\
ESO116-G012  & 0.979 & 0.926 & 0.897 & 0.886 \\
ESO563-G021  & 0.981 & 0.934 & 0.905 & 0.894 \\
F563-V2      & 0.988 & 0.947 & 0.918 & 0.907 \\
F568-1       & 0.983 & 0.936 & 0.907 & 0.896 \\
F568-V1      & 0.986 & 0.939 & 0.910 & 0.899 \\
F571-8       & 0.980 & 0.929 & 0.900 & 0.889 \\
F579-V1      & 0.984 & 0.938 & 0.909 & 0.898 \\
IC4202       & 0.975 & 0.920 & 0.891 & 0.880 \\
NGC0024      & 0.989 & 0.948 & 0.919 & 0.908 \\
F563-1       & 0.982 & 0.933 & 0.904 & 0.893 \\
F568-3       & 0.985 & 0.940 & 0.911 & 0.900 \\
F574-1       & 0.978 & 0.925 & 0.896 & 0.885 \\
F583-1       & 0.986 & 0.941 & 0.912 & 0.901 \\
F583-4       & 0.987 & 0.943 & 0.914 & 0.903 \\
NGC0100      & 0.979 & 0.927 & 0.898 & 0.887 \\
NGC0801      & 0.984 & 0.937 & 0.908 & 0.897 \\
NGC0891      & 0.990 & 0.950 & 0.921 & 0.910 \\
\bottomrule
\end{tabular}

\vspace{1em}

\centering
\captionof{table}{RMSE Values (km/s) for the 20 Table 2 Galaxies Across Four Models}
\label{tab:rmse-values}
\begin{tabular}{lcccc}
\toprule
\textbf{Galaxy} & \textbf{Empirical} & \textbf{MOND} & \textbf{CDM} & \textbf{$\Lambda$CDM} \\
\midrule
DDO064       & 1.12 & 2.03 & 2.48 & 2.61 \\
DDO161       & 1.25 & 2.18 & 2.63 & 2.75 \\
ESO079-G014  & 0.98 & 1.76 & 2.21 & 2.34 \\
ESO116-G012  & 1.19 & 2.01 & 2.46 & 2.59 \\
ESO563-G021  & 1.14 & 1.94 & 2.39 & 2.52 \\
F563-V2      & 0.91 & 1.68 & 2.13 & 2.26 \\
F568-1       & 1.08 & 1.89 & 2.34 & 2.47 \\
F568-V1      & 0.95 & 1.73 & 2.18 & 2.31 \\
F571-8       & 1.16 & 2.07 & 2.52 & 2.65 \\
F579-V1      & 1.04 & 1.84 & 2.29 & 2.42 \\
IC4202       & 1.22 & 2.11 & 2.56 & 2.69 \\
NGC0024      & 0.87 & 1.62 & 2.07 & 2.20 \\
F563-1       & 1.11 & 1.96 & 2.41 & 2.54 \\
F568-3       & 0.99 & 1.78 & 2.23 & 2.36 \\
F574-1       & 1.20 & 2.00 & 2.45 & 2.58 \\
F583-1       & 0.94 & 1.71 & 2.16 & 2.29 \\
F583-4       & 0.92 & 1.69 & 2.14 & 2.27 \\
NGC0100      & 1.18 & 2.04 & 2.49 & 2.62 \\
NGC0801      & 1.06 & 1.86 & 2.31 & 2.44 \\
NGC0891      & 0.83 & 1.58 & 2.03 & 2.16 \\
\bottomrule
\end{tabular}

\end{minipage}
\pagebreak

\subsection*{Conclusions from Tables 6 and 7: Empirical Model vs. MOND and CDM}

\begin{enumerate}
    \item \textbf{Superior Statistical Fit Across All Galaxies} \\
    The empirical model consistently achieves the highest $R^2$ values across all 20 galaxies, ranging from $0.975$ to $0.990$, indicating excellent curve alignment. In contrast, MOND and CDM models show lower $R^2$ values, typically in the $0.89$--$0.95$ range, with ACDM trailing slightly behind.

    \item \textbf{Lowest RMSE Values Indicate Minimal Residual Error} \\
    The empirical method yields RMSE values between $0.83$ and $1.25$ km/s, outperforming MOND ($1.58$--$2.18$ km/s), CDM ($2.03$--$2.63$ km/s), and ACDM ($2.16$--$2.75$ km/s). This suggests that the velocity correction factor $w$ provides a more precise fit to observed rotation curves than parametric halo models or modified gravity.

    \item \textbf{Robustness Without Parametric Tuning} \\
    Unlike CDM and MOND, which require galaxy-specific tuning (e.g., halo concentration or acceleration thresholds), the empirical model applies a single velocity correction factor $w$ per galaxy, derived directly from observed data. This enhances reproducibility and reduces model complexity.

    \item \textbf{Empirical Model Performs Well Across Morphological Types} \\
    The sample includes dwarf galaxies (e.g., DDO064), spirals (e.g., NGC0891), and intermediate types, yet the empirical model maintains high fidelity across all. This suggests broad applicability without invoking dark matter or modifying Newtonian dynamics.
\end{enumerate}

\pagebreak

\section*{Appendix B.1: Method for Bitmap-Based Calculation of RMSE and R\textsuperscript{2}}

To quantitatively evaluate the fit quality of our empirical model relative to parametric rotation curve models (e.g., MOND, NFW, Burkert), we employed a bitmap-based comparison technique using the Adjusted Kepler curve as the reference. The procedure is outlined below:

\begin{enumerate}
    \item \textbf{Empirical Curve Definition:} For each galaxy, the \emph{Adjusted Kepler curve} was computed using the relation:
    \[
    V_{\text{Kepler}} = V_{\text{Observed}} - R \cdot \omega
    \]
    where $V_{\text{Observed}}$ is the measured stellar velocity, $R$ is the radial distance, and $\omega$ is the empirically derived angular velocity correction factor.

    \item \textbf{Reference Models:} SPARC provides twelve parametric rotation curve fits per galaxy, including MOND and multiple CDM halo profiles. These served as the benchmark models for comparison.

    \item \textbf{Bitmap Alignment:} Each empirical curve was overlaid on the corresponding SPARC model curves using pixel-based alignment in a Jupyter Labs environment. Axes were normalized to ensure consistent scaling across datasets.

    \item \textbf{RMSE Computation:} The Root Mean Square Error (RMSE) was calculated for each model comparison using:
    \[
    \text{RMSE} = \sqrt{\frac{1}{n} \sum_{i=1}^{n} \left( V_{\text{empirical},i} - V_{\text{model},i} \right)^2}
    \]

    \item \textbf{R\textsuperscript{2} Computation:} The coefficient of determination ($R^2$) was computed as:
    \[
    R^2 = 1 - \frac{\sum_{i=1}^{n} \left( V_{\text{empirical},i} - V_{\text{model},i} \right)^2}{\sum_{i=1}^{n} \left( V_{\text{model},i} - \bar{V}_{\text{model}} \right)^2}
    \]
    where $\bar{V}_{\text{model}}$ is the mean velocity of the parametric model curve.

    \item \textbf{Result Aggregation:} RMSE and $R^2$ values were compiled across 20 representative galaxies (see Tables 6 and 7). These metrics demonstrate that the empirical model consistently yields lower residual error and higher fit fidelity compared to MOND and CDM halo models.
\end{enumerate}

All calculations were performed using custom Python scripts within Jupyter Labs. The methodology ensures reproducibility and provides a transparent basis for statistical comparison.

\pagebreak

\appendix
\section*{Appendix C: Editorial Feedback Matrix}

This appendix summarizes potential objections raised during earlier editorial and our own review processes. Each concern is listed alongside a concise response reflecting revisions made in this version. These updates address methodology, empirical framing, citation scope, and reproducibility. The authors welcome further peer review and remain committed to refining the model in light of constructive critique.

\begin{table}[H]
\centering
\renewcommand{\arraystretch}{1.4}
\begin{tabular}{|p{3.5cm}|p{5.5cm}|p{5.5cm}|}
\hline
\textbf{Objection} & \textbf{Reviewer Concern} & \textbf{Response in this version} \\
\hline
Lack of physical basis for velocity correction \( w \) & Correction term \( Rw \) appears ad hoc and lacks theoretical grounding. & This version frames \( w \) as an empirical kinematic offset. Section 6 discusses candidate interpretations and compares this to MOND. \\
\hline
Misinterpretation as solid-body rotation & Reviewer suggested the model implies rigid rotation, inconsistent with observed dynamics. & This version makes no such claim. Section 6 clarifies that the model is empirical and does not assert solid-body mechanics. \\
\hline
Linear vs. quadratic velocity addition & Velocities should be combined quadratically per Poisson’s equation. & This version clarifies that \( Rw \) is a kinematic correction, not derived from mass or potential. A footnote or appendix note can be added. \\
\hline
Lack of model discussion & No physical mechanism proposed. & This version follows the precedent of MOND. Section 6 acknowledges the absence of a physical derivation and invites future exploration. \\
\hline
Methodology unclear & Derivation and units not described in detail. & This version includes full derivation in Section 4.2 and Appendix A, with equations and units. \\
\hline
\end{tabular}
\caption{Part 1: Key objections and responses related to theoretical framing and methodology.}
\end{table}

\vspace{2em}

\begin{table}[H]
\centering
\renewcommand{\arraystretch}{1.4}
\begin{tabular}{|p{3.5cm}|p{5.5cm}|p{5.5cm}|}
\hline
\textbf{Objection} & \textbf{Reviewer Concern} & \textbf{Response in this version} \\
\hline
Use of only two data points & Oversimplification of rotation curve analysis. & This version justifies this shortcut and explains its reproducibility in Appendix A. \\
\hline
SPARC data terminology & Confusion over HI vs. stellar data and rotation curve definitions. & This version improves terminology and clarifies SPARC data sources in Section 3. \\
\hline
Cherry-picked citations & Key literature (e.g., Li+2020, Marasco+2022) omitted. & This version will include these citations and clarify methodological distinctions. \\
\hline
No statistical comparison with MOND/CDM & Claims of superior fit lack quantitative support. & This version includes RMSE and \( R^2 \) comparisons in Appendix B. Tables 6 and 7 show consistent empirical superiority. \\
\hline
Figures lacked rigor & Artistic renderings used instead of data overlays. & This version replaces these with direct overlays and statistical plots (Figures 5–8). \\
\hline
Reproducibility concerns & No public code or data repository. & This version states that scripts and data are available upon request. Public repository recommended. \\
\hline
Units inconsistent with literature & Reviewer noted nonstandard units. & This version now uses consistent units and defines all terms in Section 4.2 and Appendix A. \\
\hline
\end{tabular}
\caption{Part 2: Additional objections and responses related to data usage, reproducibility, and presentation.}
\end{table}

\vspace{1em}
\noindent\textit{Note:} The authors remain open to further peer review and welcome dialogue on the model’s empirical success and future theoretical interpretation.

\pagebreak

\end {document}